\documentclass[longbibliography,physrev,aps,superscriptaddress,twocolumn,nofootinbib]{revtex4-2}

\usepackage[utf8]{inputenc}
\usepackage{amsmath}
\usepackage{mathtools}
\usepackage{bm}
\usepackage[colorlinks=true,citecolor=blue,linkcolor=blue,urlcolor=blue]{hyperref}
\usepackage{xcolor}
\usepackage{bbold}
\usepackage{float}
\usepackage{braket}
\usepackage{orcidlink}
\usepackage{lipsum}
\usepackage{transparent}

\usepackage[german,english]{babel}

\makeatletter
\def\bbl@set@language#1{%
	\edef\languagename{%
		\ifnum\escapechar=\expandafter`\string#1\@empty
		\else\string#1\@empty\fi}%
	\@ifundefined{babel@language@alias@\languagename}{}{%
		\edef\languagename{\@nameuse{babel@language@alias@\languagename}}%
	}%
	\select@language{\languagename}%
	\expandafter\ifx\csname date\languagename\endcsname\relax\else
	\if@filesw
	\protected@write\@auxout{}{\string\select@language{\languagename}}%
	\bbl@for\bbl@tempa\BabelContentsFiles{%
		\addtocontents{\bbl@tempa}{\xstring\select@language{\languagename}}}%
	\bbl@usehooks{write}{}%
	\fi
	\fi}
\newcommand{\DeclareLanguageAlias}[2]{%
	\global\@namedef{babel@language@alias@#1}{#2}%
}
\makeatother

\DeclareLanguageAlias{en}{english}
\DeclareLanguageAlias{English}{english}
\DeclareLanguageAlias{Englisch}{english}
\DeclareLanguageAlias{EN}{english}
\DeclareLanguageAlias{en-US}{english}
\DeclareLanguageAlias{de}{german}

\usepackage{graphicx}
\usepackage[export]{adjustbox}
\usepackage[capitalize]{cleveref}
\newcommand{\ham}{H}

\begin{document}
	\title{Graph-theoretical approach to the eigenvalue spectrum of perturbed higher-order exceptional points}

     \author{Daniel Grom\,\orcidlink{0009-0003-7407-378X}}
     \email{daniel.grom@ovgu.de}
    \affiliation{%
		Institut für Physik, Otto-von-Guericke-Universität Magdeburg, Postfach 4120, D-39016 Magdeburg, Germany
	}%

     \author{Julius Kullig\,\orcidlink{0000-0002-8137-9408}}
     \email{julius.kullig@ovgu.de}
    \affiliation{%
		Institut für Physik, Otto-von-Guericke-Universität Magdeburg, Postfach 4120, D-39016 Magdeburg, Germany
	}%
 
	\author{Malte Röntgen\,\orcidlink{0000-0001-7784-8104}}
 \email{malte.rontgen@univ-lemans.fr}
\affiliation{%
	Laboratoire d’Acoustique de l’Université du Mans, Unite Mixte de Recherche 6613, Centre National de la Recherche Scientifique, Avenue O. Messiaen, F-72085 Le Mans Cedex 9, France
}%

	\author{Jan Wiersig\,\orcidlink{0000-0003-2764-1660}}%
    \email{jan.wiersig@ovgu.de}
	    \affiliation{%
		Institut für Physik, Otto-von-Guericke-Universität Magdeburg, Postfach 4120, D-39016 Magdeburg, Germany
	}%

\begin{abstract}
    Exceptional points are special degeneracy points in parameter space that can arise in (effective) non-Hermitian Hamiltonians describing open quantum and wave systems. At an $n$-th order exceptional point, $n$ eigenvalues and the corresponding eigenvectors simultaneously coalesce. 
    These coalescing eigenvalues typically exhibit a strong response to small perturbations which can be useful for sensor applications. A so-called generic perturbation with strength $\epsilon$ changes the eigenvalues proportional to the $n$-th root of $\epsilon$. A different eigenvalue behavior under perturbation is called non-generic. 
    An understanding of the behavior of the eigenvalues for various types of perturbations is desirable and also crucial for applications. We advocate a graph-theoretical perspective that contributes to the understanding of perturbative effects on the eigenvalue spectrum of higher-order exceptional points, i.e. $n>2$. To highlight the relevance of non-generic perturbations and to give an interpretation for their occurrence, we consider an illustrative example, a system of microrings coupled by a semi-infinite waveguide with an end mirror. Furthermore, the saturation effect occurring for cavity-selective sensing in such a system is naturally explained within the graph-theoretical picture. 
\end{abstract}
	
\maketitle

\section{Introduction}	

Real world physical systems interact with their environment, which leads, among other effects, to a dissipative nature of the system. In first approximation, these interactions are often neglected, which can lead to a suppression of intriguing phenomena. The field of non-Hermitian physics aims to uncover and describe such effects in open quantum and wave systems~\cite{Ashida2020}. 

An especially interesting feature of non-Hermitian systems is given by the occurrence of exceptional points (EPs) in parameter space, where $n$ eigenvalues and the corresponding eigenvectors of the (effective) Hamiltonian coalesce~\cite{Kato66,WBW96,Heiss00,Berry04,Heiss04}. Second-order EPs (that is, $n=2$) have been explored in many different physical systems, especially in photonics~\cite{MA19}. The physics and the potential usage of higher-order EPs with $n>2$ are by far less explored. 
EPs of third order have been experimentally realized recently using coupled acoustic~\cite{DMX16}, photonic cavities~\cite{HHW17}, and high-index dielectric spheres in the microwave regime~\cite{WHL19}. Potential applications of higher-order EPs are on-chip integrated microlasers based on higher-order EPs~\cite{LiaZhoDu2023}, EP-based sensors~\cite{HHW17,2019_zhicheng}, flat-top optical filters~\cite{HablerS20}, and the possibility of speeding up entanglement generation~\cite{LCA23}.

In general, EP-based sensing applications~\cite{Wiersig14b,Wiersig16,COZ17} try to exploit the fact that a small perturbation of strength~$\epsilon$ leads to an eigenvalue response stronger than a linear $\epsilon$-scaling of conventional sensors.
A perturbation of an $n$-th-order EP leading to an eigenvalue splitting proportional to $\epsilon^{1/n}$ is called generic~\cite{Kato66}.

However, there is no guarantee that a perturbation is generic for a given EP; see e.g. Refs.~\cite{1998_Ma,2008_Graefe,2012_demange,XZH19,Wiersig22}. Non-generic perturbations are utilized, for instance, for the concept of exceptional surfaces~\cite{ZhoRenKha2019}. Here, specific parameter changes corresponding to fabrication tolerances leave the system at the EP, while other parameter changes correspond to the perturbations that are to be detected. In this way, the system is at the same time immune to fabrication tolerances and exhibits a strong response to certain perturbations. 
Further examples for systems with non-generic perturbations are the following.
Reference~\cite{2019_zhicheng} reports a parity-time ($\mathcal{PT}$)-symmetric electronic circuit exhibiting a sixth-order EP.
Here, a non-generic perturbation is present where the eigenvalue splitting is proportional to $\epsilon^{1/4}$. 
Another example are waveguide-coupled microrings with asymmetric backscattering~\cite{2023_kullig}, where generic and different kinds of non-generic perturbations can be introduced by placing a small scatterer near different microrings. These easy-to-realize non-generic perturbations are good examples for the necessity of understanding why and under which conditions non-generic perturbations are present. 
Recently, Y.-X. Xiao \textit{et al.} tackled this question in Ref.~\cite{2023_xiao} by transforming the Hamiltonian to the Jordan-normal form and applying certain splitting rules. The disadvantage of their approach is that a transformation to the Jordan-normal form is numerically unstable and an intuitive understanding of the system's behavior in the new basis is not possible. 

The aim of this paper is to introduce a graph-theoretical approach to the eigenvalue spectrum of perturbed higher-order EPs. 
Our work shifts the complexity of finding a Jordan-normal transformation into finding graph-theoretical entities for the characterization of the system's behavior under perturbation. 
A crucial advantage of this approach is that it is independent of the basis used for the effective Hamiltonian.
Hence, one can freely work in a basis which is suited for an intuitive physical interpretation. 
In order to carry out the graph-theoretical approach, methods from the field of combinatorics, especially the Coates determinant formula~\cite{Brualdi2008CombinatorialApproachMatrixTheory}, are combined with an algorithm from the field of plane algebraic curves to determine the Puiseux series based on the Newton polygon~\cite{1950_walker}.
As a result, we are able to express the asymptotically leading order terms of the Puiseux series, which characterize the perturbed eigenvalue spectrum, with graph-theoretical entities. 

The structure of the paper is as follows. The main text describes the approach and familiarizes the reader with the utilized concepts, whereas the appendix contains the technical details.
We start with an introduction to graph theory in \cref{sec:graph_foundation}.
\cref{sec:puiseux_main} briefly discusses the eigenvalue spectrum of an (effective) Hamiltonian. The Puiseux series is introduced and connected to the eigenvalue spectrum in \cref{sec:puiseux_foundation}. 
From there on the graph-theoretical approach is presented in \cref{sec:graph_approach}. 
In \cref{sec:physical_examples}, the approach is applied to give an interpretation for the occurrence of different eigenvalue behaviors for perturbed microring systems. 

\section{Graph-theoretical foundation} \label{sec:graph_foundation}
\begin{figure}[tb]
	\begin{center}
		\includegraphics[width=1.0\columnwidth]{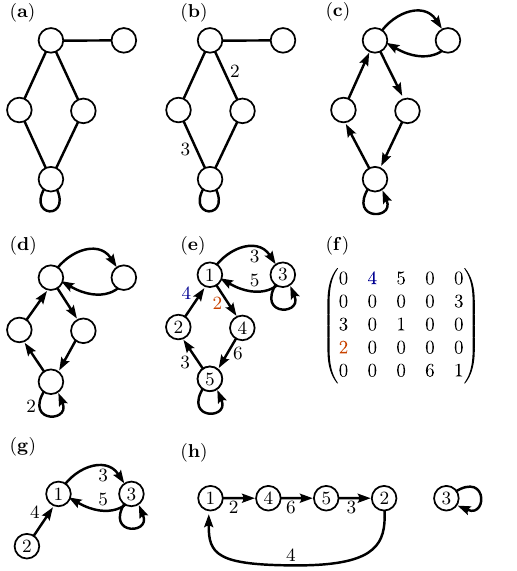}
		\caption{(a) - (d) Different graphs. (a) Unweighted (all edges having a weight of unity), undirected (no edges with arrows). (b) The same graph as in (a), but with additional weights on two edges, rendering the graph weighted and undirected. (c) Same graph as in (a), but giving a direction to the edges, thus making the graph unweighted and directed. (d) Giving a weight to one of the edges of (c) results in a weighted and directed graph. (e) Another weighted directed graph, whose matrix representation is shown in (f). (g) shows the induced subgraph of (e) on the vertex set $\{1,2,3\}$. (h) shows the only linear subdigraph of (e). In all graphs, any edge without a number has a weight of unity.
        }
		\label{fig:GraphTheoryIntro}
	\end{center}
\end{figure}

Before we introduce the two graph-theoretical core concepts of our method, let us equip the reader with some basic knowledge on graph theory and its applications.
In layman's terms, a graph is nothing but a network of nodes (called vertices) that are connected by links (called edges).
An example graph consisting of five vertices is depicted in \cref{fig:GraphTheoryIntro}(a).
In this graph, all of the six edges are unweighted and undirected.
In \cref{fig:GraphTheoryIntro}(b)-(d), other example graphs are shown whose edges are weighted and directional, or both.
What makes graph theory so powerful is its versatility.
Indeed, many artificial and real networks can be described by graphs, with examples including transportation networks (roads, for instance), the Internet, social networks, electric circuits, and many more \cite{Newman2018NetworksIntroduction}.
In complete analogy to these numerous examples, algorithms based on graph theory permeate our daily lives.
For instance, Dijkstra's algorithm \cite{Dijkstra1959NM1269NoteTwoProblemsConnexion} is a classic method of finding the shortest path between two vertices of a network \footnote{We remark that Dijkstra's algorithm only works in graphs with non-negative weights; for graphs with such negative weights, one can use the Bellmann-Ford algorithm \cite{Ford1956NetworkFlowTheory,Bellman1958QAM1687RoutingProblem}.}; in more specialized forms, it is the basis of modern car routing systems \footnote{One such more specialized form that makes use of the hierarchical character of road networks is the contraction hierarchies algorithm; see for instance Ref. \cite{Geisberger2008EA319ContractionHierarchiesFasterSimpler}.} and the Internet \footnote{See, for instance, Chapter 10 of Ref. \cite{Dijkstra1959NM1269NoteTwoProblemsConnexion}}.
On another page, the need for judging the importance of individual nodes in a network lead to various graph-theoretical measures; one of them, the so-called eigenvector centrality, is the basis of Google's famous PageRank algorithm.
Graph theory also has applications in fundamental science, some examples being molecular graph theory \cite{Balaban1985JCICS25334ApplicationsGraphTheoryChemistry,Tsuji2018CR1184887QuantumInterferenceGraphsWalks}, finding hidden symmetries in diverse physical setups \cite{Smith2019PA514855HiddenSymmetriesRealTheoretical,Rontgen2021GraphtheoreticalAnalysisLocalLatent,Rontgen2022LatentSymmetriesIntroduction}, or, very recently, the treatment of exceptional points in scattering~\cite{Scali2023JPAMT56275201GraphTheoryApproachExceptional}.
Before we continue, we refer the interested reader to Ref.~\cite{Newman2018NetworksIntroduction}, which provides an application-oriented introduction to modern graph theory.

In the context of this work, we are mainly interested in graph theory due to a convenient one-to-one mapping \cite{Newman2018NetworksIntroduction}: Given an $n\times n$ matrix $A$, this matrix can be uniquely identified with a graph that consists of $n$ vertices, with each matrix element $A_{i,j}$ corresponding to a (possibly directed) edge between vertices $j$ and $i$.
This mapping, which is visualized in \cref{fig:GraphTheoryIntro}(e) and (f), allows us to use graph-theoretical concepts in the treatment of our problem.
We remark that this mapping is, generally speaking, very useful in tight-binding models and models emerging from coupled-mode theory.

There are two important graph-theoretical concepts that we shall use in this work.
The first concept is that of an \emph{induced subgraph}.
Given a graph $G$ and some of its vertices $S$, the induced subgraph $G\left[S\right]$ is obtained from~$G$ by keeping only the vertices $S$ and all the edges between them; this includes loops, that is, edges connecting a vertex to itself.
In \cref{fig:GraphTheoryIntro}(g), we show the induced subgraph of the graph depicted in subfigure (e), with the choice $S=\{1,2,3\}$.
The relevance of this concept for our approach arises from the calculation of the characteristic polynomial in a combinatorial manner \cite{1990Marcus, Brualdi2008CombinatorialApproachMatrixTheory}. We will pick up on this in \cref{sec:graph_quantities}.

The second concept that we shall use is that of a \emph{linear subdigraph} (LSD)~\cite{Brualdi2008CombinatorialApproachMatrixTheory}.
Given a graph $G$, a LSD is obtained from $G$ by deleting any number of edges (but keeping all vertices) such that, after this process, any vertex possesses exactly one incoming and one outgoing edge \footnote{For this purpose, note that one treats an undirected edge between two vertices $i,j$ as two separate directed edges; one from $i$ to $j$, and another one from $j$ to $i$.}.
Loops, that is, edges connecting a vertex to itself, are thereby counted both as one incoming and one outgoing edge.
The resulting graph is thus (i) a subgraph of $G$, (ii) directed, and could also be drawn along a line, thus the naming.
There may be different LSDs corresponding to a given graph; for \cref{fig:GraphTheoryIntro}(e), however, there is only one, which we depict in \cref{fig:GraphTheoryIntro}(h).
We note that this graph consists of two disconnected components---in each of which the edges form cycles---; in graph theory, a graph may contain several disconnected networks.
Before we continue, we remark that other applications of LSDs are described in chapter 10 of Ref.~\cite{Brualdi2008CombinatorialApproachMatrixTheory}.

The so-called Coates determinant formula~\cite{Brualdi2008CombinatorialApproachMatrixTheory,Coates1959ITCT6170FlowGraphSolutionsLinearAlgebraic},
\begin{equation}
    \det{A} = \sum_{L \in \mathcal{L} (A)} (-1)^{n - c (L)} w (L) \,,
	\label{eq:coates_det_main}
\end{equation}
plays a central role in Ref.~\cite{Scali2023JPAMT56275201GraphTheoryApproachExceptional}, as well as in this work, where we use it to determine the coefficients of the characteristic polynomial; we shall discuss this in \cref{sec:graph_quantities}.
Conceptually speaking, the Coates determinant formula links the determinant of an $n\times{}n$ matrix $A$ to the set $\mathcal{L} (A)$  of its LSDs.
For a given LSD $L$ in this set, the number of disconnected components is denoted by $c(L)$ and the product of its edge weights by $w(L)$. 

As an instructive example, let us now apply the Coates determinant formula to the graph depicted in \cref{fig:GraphTheoryIntro}(e).
Direct calculation gives us $-144$ for the left-hand side of \cref{eq:coates_det_main}.
For the right-hand side, we see that the sum collapses to a single term, as the LSD depicted in \cref{fig:GraphTheoryIntro}(h) is the only one.
It consists of $c=2$ disconnected components, so that the prefactor $(-1)^{n - c (L)} = -1$; the product of weights is $2 \cdot 6 \cdot 3 \cdot 4 \cdot 1 = 144$.
Thus, the right-hand side evaluates to $-144$, as expected.

\section{Spectrum of the effective Hamiltonian} \label{sec:puiseux_main}

After the above introduction to graph theory, let us now go into the details of the eigenvalue problem which we want to treat with our graph-theoretical approach.
We consider an effective $n\times n$ Hamiltonian of the form
\begin{equation}
	H(\epsilon) = \underbrace{H_0}_{\text{nilpotent}} +\hspace{.35cm} \epsilon \, H_1 \,.
	\label{eq:effective_hamiltonian}
\end{equation}
For convenience, dimensionless units are used. 
The above Eq.~\eqref{eq:effective_hamiltonian} includes the case where~$H_0$ exhibit an EP of order~$n$. Additionally, it contains the situation where multiple EPs of different orders but the same degenerate eigenvalues are present. 
Without loss of generality, we assume in the following that all of these eigenvalues are equal to zero, which can be always achieved with a simple global shift of the spectrum of $H_0$.
The cases above can be formulated as $H_0$ being nilpotent of order~$n$, which means~$H_0^n = 0$ but $H_0^{n-1} \neq 0$, or consisting of nilpotent blocks with lower order.
A linear perturbation described by~$H_1$ and the perturbation parameter $\epsilon$ is added to $H_0$ to form the Hamiltonian of interest for our purpose. The structure and the concrete values of the matrix elements of~$H_1$ determine the characterization of the perturbation as generic or non-generic, as we will discuss with the help of LSDs in \cref{sec:graph_approach}.  

The right and left eigenvalue problem of the full Hamiltonian in Eq.~(\ref{eq:effective_hamiltonian}) read as 
\begin{eqnarray}
	H | \psi^R \rangle & = & \lambda | \psi^R \rangle \label{eq:eig_problem} \\
	\langle \psi^L | H & = & \langle \psi^L | \lambda,
\end{eqnarray}
where $\lambda \in \mathbb{C}$ is the eigenvalue, $| \psi^R \rangle$ the right eigenvector, and $\langle \psi^L |$ the left eigenvector. Due to the non-Hermiticity of $H$, the left and right eigenvectors are not necessarily the adjoint of each other. 

For a given perturbation strength $\epsilon$, the eigenvalues of~$H(\epsilon)$ are the roots $\lambda$ of the characteristic equation
\begin{equation}
    P_{\text{char}}(\lambda, \epsilon) = 0 \,,
    \label{eq:char_eq}
\end{equation}
with $P_{\text{char}}(\lambda, \epsilon) := \det{[\lambda \mathbb{1} - H(\epsilon)]}$ being the characteristic polynomial of $H$. We use this definition since it renders the characteristic polynomial to be monic, that is, the prefactor of $\lambda^n$ is equal to unity.
Importantly, the characteristic polynomial is independent of the basis of the Hamiltonian \cite{2024_Axler}.
Therefore, it is a reasonable starting point for our goal to formulate a basis-independent graph-theoretical approach to the eigenvalue spectrum. 

\section{The Puiseux series} \label{sec:puiseux_foundation}

For our graph-theoretical approach,
we shall use techniques from two different branches of mathematics, namely, from graph theory---which was introduced in \cref{sec:graph_foundation}---and from the theory of algebraic curves.
To see the connection of our setup to the latter, let us go back to the characteristic \cref{eq:char_eq}.
Instead of using this equation to find the eigenvalues $\lambda$ of $\ham$ for a specific $\epsilon$, one could also investigate the set of all possible pairs $(\epsilon, \lambda)$ for which \cref{eq:char_eq} is fulfilled.
In a four-dimensional coordinate system (whose axes are the real and imaginary parts of $\epsilon,\lambda$), this point set forms a so-called plane algebraic curve.
We note that so-called eigenvalue braids and knots arising by encircling a single EP \cite{2022_patil} or multiple EPs \cite{2024_guria} in parameter space, which are recently investigated, can be described in the context of plane algebraic curves \cite{2012_brieskorn_alg_curves}.
An introduction to the topic of algebraic curves can be found, for instance, in Ref.~\cite{1950_walker}.

We focus on one specific result of the theory of algebraic curves.
Namely, we will parametrize $\lambda$ by $\epsilon$; this results in a Puiseux series in fractional powers. 
The structure of this series is
\begin{equation}
    \lambda (\epsilon ) = \sum_{l=0}^{\infty} p_l \epsilon^{\frac{l}{k}} ,
    \label{eq:puiseux_struc}
\end{equation}
where $p_l \in \mathbb{C}$ and $k \in \mathbb{N}$ \cite{1994_fischer_alg_kurven}.
\Cref{eq:puiseux_struc} is a convenient and powerful tool to obtain the eigenvalue spectrum of $\ham(\epsilon)$, since it gives us the spectrum without the need to (numerically) diagonalize $\ham$ for different perturbation strengths $\epsilon$.
This advantage becomes even more elevated for small enough perturbation strengths, where the first non-vanishing (that is, leading-order) term of \cref{eq:puiseux_struc} becomes dominant.
Since the characteristic \cref{eq:char_eq} contains at maximum $n$ unique solutions for the eigenvalues $\lambda$, the same amount of unique Puiseux series for a given characteristic equation exist; we will refer to them as eigenvalue branches.
Concretely, in leading order the Puiseux series of the $i$-th eigenvalue branch reads as
\begin{equation}
    \lambda_i (\epsilon ) = p_{\alpha}^{(i)} \epsilon^{\kappa_{\alpha}^{(i)}} + \cdots \, ,
    \label{eq:puiseux_leading}
\end{equation}
with the index $\alpha$ indicating the fact that we are looking at the leading-order term.

In practice, many eigenvalue branches (sometimes all) share the same fractional power, and only differ (if at all) in the coefficient.
Thus, it makes sense to take this fact into account by rewriting \cref{eq:puiseux_leading} in a more structured form, that is,
\begin{equation}
\lambda_m^{(k)} (\epsilon) = (p_{\alpha}^{(k)})_m \epsilon^{\kappa_{\alpha}^{(k)}} \,,
\label{eq:branches-k-segment}
\end{equation}
where the index $k$ denotes what we call here the ``main eigenvalue branch'', and the index $m$ denotes the ``minor eigenvalue branch''.
We remark that not all main eigenvalue branches need to have the same amount of minor eigenvalue branches.

The fractional power of the $k$-th main branch, as well as the individual coefficients $(p_{\alpha}^{(k)})_m$ of its minor eigenvalue branches can be found by an algorithm based on Newton polygons \cite{1950_walker}, which is summarized in \cref{app:plane_alg_curves}, and which we build upon in our graph-theoretical approach in the next \cref{sec:graph_approach}.

Before we introduce our method, let us apply the above introduced concepts to a specific example.
We consider two waveguide-coupled microrings with a mirror-induced asymmetric backscattering exhibiting an EP of order~4~\cite{2023_kullig}. 
At the EP, the unperturbed system is described by
\begin{equation}
    H_{0}^{\text{EP}_4} = \left(
\begin{array}{c c c c}
0 & A_{1,2} & 0 & 0 \\
0 & 0 & A_{2,3} & 0 \\
0 & 0 & 0 & A_{3,4} \\
0 & 0 & 0 & 0 \\
\end{array}
    \right) \,.
    \label{eq:H0_example_2_ring}
\end{equation}
Placing a small particle in the vicinity of the microring furthest away from the mirror leads to a generic perturbation that can be modeled through
\begin{equation}
    H_{1}^{\text{gen}} = \left(
\begin{array}{c c c c}
e_{1,1} & 0 & 0 & e_{1,4} \\
0 & 0 & 0 & 0 \\
0 & 0 & 0 & 0 \\
e_{4,1} & 0 & 0 & e_{4,4} \\
\end{array}
    \right) .
    \label{eq:H1_example_2_ring_generic}
\end{equation}
The eigenvalue spectrum of the Hamiltonian
\begin{equation}
H_{\text{gen}}^{\text{EP}_4} = H_{0}^{\text{EP}_4} + \epsilon H_{1}^{\text{gen}} \label{eq:H_example_2ring}
\end{equation}
is obtained with the related Puiseux series in leading order
\begin{equation}
    \lambda_{m}^{\text{gen}} = (A_{1,2} A_{2,3} A_{3,4} e_{4,1})^{\frac{1}{4}} e^{\frac{1}{2} \pi i m} \epsilon^{\frac{1}{4}} + \ldots 
    \label{eq:puiseux_example_generic}
\end{equation}
for small perturbations.
The index $m = 1, ...,4$ gives rise to the different eigenvalue branches.
A comparison of \cref{eq:puiseux_example_generic,eq:branches-k-segment} reveals the dominant coefficients $(p_{\alpha})_m = (A_{1,2} A_{2,3} A_{3,4} e_{4,1})^{\frac{1}{4}} e^{\frac{1}{2} \pi i m}$ to the same dominant fractional power $\kappa_{\alpha} = 1/4$.
Following our classification below \cref{eq:branches-k-segment}, we see that there is only one main branch with four minor branches.
\begin{figure}[]
	\begin{center}
		\includegraphics[width=1.0\columnwidth]{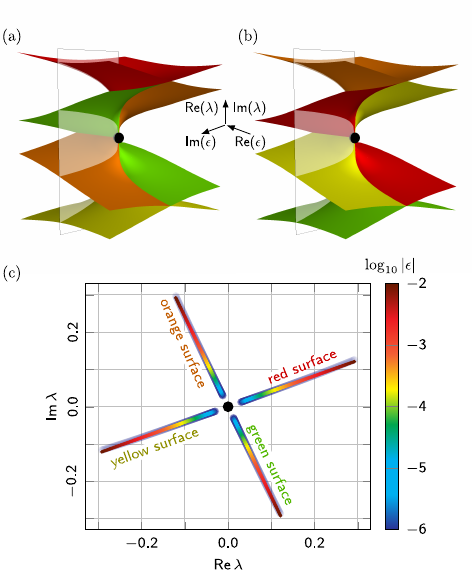} 
		\caption{The real and imaginary part of the dimensionless eigenvalue spectrum obtained by the Puiseux series in~\cref{eq:puiseux_example_generic} for the system described by the Hamiltonian in \cref{eq:H_example_2ring}, where all parameters appearing in the matrix are set to one, over the complex perturbation parameter $\epsilon$ in (a) and (b) respectively. The colours correspond to different branches from the Puiseux expansion. In (c) the complex eigenvalue plane is shown. The data is given for a pure positive imaginary part of $\epsilon$, ranging from $(i \cdot 10^{-6})$ to $(i \cdot 10^{-2})$,  see the cut plane in (a) and (b). The colour map shows the absolute value of the perturbation strength in logarithmic scale. The annotations refer to the surfaces from (a) and (b). 
			The numeric solutions to the eigenvalue problem in Eq.~\eqref{eq:eig_problem} are indicated with the shadowed branches in the background.
		}
		\label{fig:riemann_generic}
	\end{center}
\end{figure}

In \cref{sec:leading_puiseux}, we show how the leading-order term of \cref{eq:puiseux_example_generic} can be found using our graph-theoretical approach. 
\Cref{fig:riemann_generic} shows the eigenvalues of \cref{eq:H_example_2ring} as a function of the complex perturbation parameter~$\epsilon$, where all matrix elements $e_{i,j}$ and $A_{i,j}$ are set to unity. 
The eigenvalue branches $m = 1, \ldots, 4$ in Eq.~\eqref{eq:puiseux_example_generic} have different surface colors in Fig.~\ref{fig:riemann_generic} (a) and (b). 
The $z$-axis of subfigures (a) and (b) corresponds to the real- and imaginary part of the eigenvalues, respectively. The $x$- and $y$-coordinates correspond to the real and imaginary part of the perturbation parameter~$\epsilon$. The EP is highlighted with a black sphere at the coordinate center.

An alternative visualization of the eigenvalue spectrum is given in \cref{fig:riemann_generic}(c). From that figure, the terminology eigenvalue branches becomes evident. 
By choosing a cut plane orthogonal to the $\epsilon$-plane and touching the EP at the center of the coordinate system, one can map the intersection points from the surfaces from (a) and (b) and the cut plane to the complex eigenvalue plane. Figure~\ref{fig:riemann_generic}(c) visualizes the eigenvalues in the complex plane arising from the cut plane from (a) and (b).   
The colormap of subfigure (c) encodes the absolute perturbation strength $\epsilon$ in logarithmic scale. 
Even small perturbations have a considerable effect on the eigenvalues, due to the dominant fractional power $\kappa_{\alpha} = 1/4$ in $\epsilon$ \cite{Wiersig14b,COZ17,HHW17}.
The shadows under the eigenvalue branches in Fig.~\ref{fig:riemann_generic}(c) visualize the numerical obtained solutions of Eq.~\eqref{eq:eig_problem}. These solutions are in good agreement with the Puiseux series for the visualized parameter range of the perturbation strength~$\epsilon$. 

After this introductory example, we proceed with the main result of this paper, namely, our graph-theoretical algorithm for determining the eigenvalue branches in leading-order.

\section{Graph-theoretical approach} \label{sec:graph_approach}
Our approach evolves around expressing the coefficients of the characteristic polynomial with graph-theoretical tools and connecting these coefficients with the leading-order terms of the Puiseux series or, in other words, the eigenvalue branches in Eq.~\eqref{eq:branches-k-segment}. 

First, we start with a description of how to find the leading-order terms of the Puiseux series with the help of the Newton polygon \cite{1950_walker}, in \cref{sec:newton_polygon}.
In the context of EPs, the Newton polygon is discussed in Refs.~\cite{2023_Jaiswal,2023_banerjee}.
We proceed, in \cref{sec:searching_seq}, to point out which coefficients of the characteristic polynomial are needed to calculate the eigenvalue branches in leading order. 
The graph-theoretical aspect of the method is discussed in \cref{sec:graph_quantities} with the help of a combinatorial formula for calculating the coefficients of the characteristic polynomial independently.
Finally, \cref{sec:leading_puiseux} combines the different concepts.

\subsection{Finding the leading-order terms of the Puiseux series} \label{sec:newton_polygon}

It is a powerful result of Ref.~\cite{1950_walker} that the dominant fractional powers of the Puiseux series $\kappa_{\alpha}^{(i)}$ [Eq.~\eqref{eq:branches-k-segment}] can be determined with the help of the so-called Newton polygon.
The foundation for the construction of this polygon is the carrier 
\begin{equation}
    \Delta P_{\text{char}} = \left\{ (\mu, \nu) \in \mathbb{N}^2 | b_{\mu, \nu} \neq 0 \right\} 
    \label{eq:carrier_char_poly}
\end{equation}
of the characteristic polynomial
\begin{equation}
    P_{\text{char}}(\lambda, \epsilon) = \sum_{\nu =0}^n \sum_{\mu = 0}^n b_{\mu, \nu} \epsilon^{\mu} \lambda^{n - \nu} \,.
    \label{eq:generalcP}
\end{equation}
The carrier describes the set of index-tuples of the non-vanishing coefficients~$b_{\mu, \nu}$ of the characteristic polynomial. 
We can thus conveniently draw the carrier in a two-dimensional point grid of size $(n+1)\times (n+1)$, with the indices $\mu,\nu$ forming the coordinate axes, and with each index-tuple $(\mu,\nu)$ being an intersection point on the grid.
In the following, we shall denote these points as ``carrier points''.

\begin{figure}[tb]
	\begin{center}
		\includegraphics[width=1.0\columnwidth]{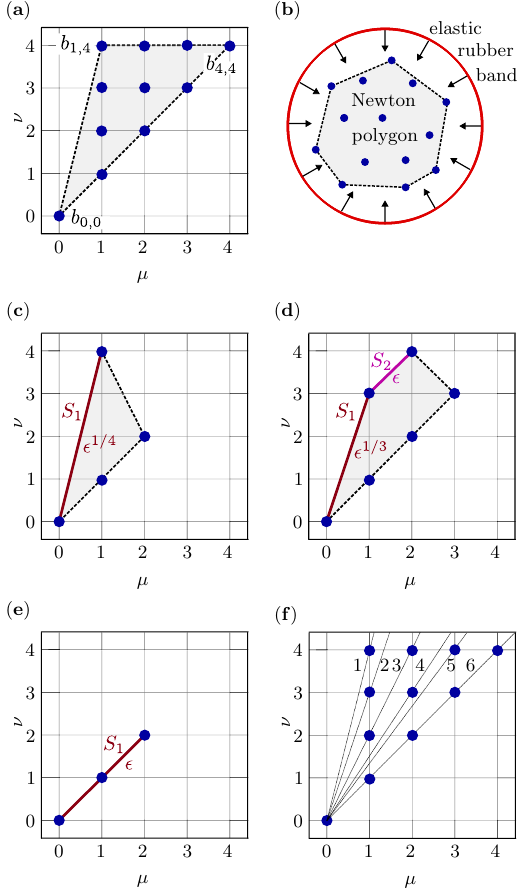}
		\caption{(a) The carrier of a generic characteristic polynomial of the form in \cref{eq:cPs}, with all $b_{\mu,\nu}$ non-vanishing. The dashed triangle denotes the Newton polygon. (b) The rubber-band construction principle of the Newton polygon. (c) to (e): Different carriers, along with their respective Newton polygons and all relevant line segments (see text for details). (f) Optimal order for evaluating the coefficients $b_{\mu,\nu}$ (see \cref{sec:searching_seq} for details).
  }
		\label{fig:carrierAndPolygon}
	\end{center}
\end{figure}

For a general polynomial [\cref{eq:generalcP}], there are no constraints on the $b_{\mu,\nu}$, and thus the carrier could comprise all $(n+1)^2$ grid points.
In this work, however, we are not dealing with a general polynomial, but with the one related to the Hamiltonian $H$ of \cref{eq:effective_hamiltonian}.
In particular, since we demand all eigenvalues of $H_0$ to be zero (see also \cref{app:char_poly}), and due to \cref{eq:effective_hamiltonian}, the characteristic polynomial has the structure
\begin{equation}
    P_{\text{char}}(\lambda, \epsilon) = \lambda^n + \sum_{\nu =1}^n \sum_{\mu = 1}^\nu b_{\mu, \nu} \epsilon^{\mu} \lambda^{n - \nu} \,.
    \label{eq:cPs}
\end{equation}
From a comparison with \cref{eq:generalcP}, we see that the coefficients $b_{\mu,\nu}$ with $\mu > \nu$ are zero; the same applies to those coefficients $b_{\mu,\nu}$ where either $\mu$ or $\nu$ vanish.
Thus, going back to \cref{eq:carrier_char_poly}, we see that the possible carrier points on our coordinate grid lie within a triangle whose corners are $(0,0)$, $(1,n)$, and $(n,n)$.
For $n=4$, this triangle as well as all possible points are shown in \cref{fig:carrierAndPolygon}(a).

We are now ready to introduce the Newton polygon.
Mathematically speaking, it is the convex hull of the carrier.
In practice, the Newton polygon can be obtained by an easy procedure: Imagine the carrier points as pinpoints, and then tightly wrap a rubber-band around it; this procedure is visualized in \cref{fig:carrierAndPolygon}(b).
The result is a polygon that consists of line segments that start and end at carrier points.
To acquaint the reader with this concept, in \cref{fig:carrierAndPolygon}(c)--(e), we show different carriers and their corresponding Newton polygons.

In practice, since our aim is to determine the Puiseux series in leading order in Eq.~\eqref{eq:branches-k-segment}, we are interested only in certain line segments of the Newton polygon.
These \emph{relevant} line segments can be obtained graphically by starting at the point $(0,0)$ and traversing the edges of the Newton polygon in clockwise direction until one hits an edge with vanishing or negative slope; the chain of $l$ edges traversed are the relevant line segments $\mathcal{S}_1,\ldots{},\mathcal{S}_l$.
We note that a given Newton polygon has at least one relevant line segment.
In \cref{fig:carrierAndPolygon}(c) to~(e), different Newton polygons along with all of their relevant line segments are shown.

As shown in Ref.~\cite{1950_walker}, the inverse slope of the $k$-th relevant line segment determines the leading-order fractional power 
\begin{equation}
    \kappa_{\alpha}^{(k)} = \frac{\mu_k - \mu_{k-1}}{\nu_k - \nu_{k-1}} 
    \label{eq:kappa_alpha_i}
\end{equation}
of the $k$-th main eigenvalue branch.
Here $(\mu_k,\nu_k)$ and $(\mu_{k-1},\nu_{k-1})$ denote the end and starting point, respectively, of the $k$-th relevant line segment; we set $(\mu_0, \nu_0)$ to $(0, 0)$.
In \cref{fig:carrierAndPolygon}(c)--(e), we have indicated the dominant eigenvalue behavior with respect to the perturbation strength~$\epsilon$ resulting from the inverse slopes of the relevant line segments.
The carrier corresponding to the example system described by the Hamiltonian in Eq.~\eqref{eq:H_example_2ring} is shown in \cref{fig:carrierAndPolygon}(c) and delivers the same $\epsilon^{1/4}$-behavior as stated in Eq.~\eqref{eq:puiseux_example_generic}.

While the leading-order fractional power of the $k$-th main eigenvalue branch is thus directly visible in the Newton polygon, the corresponding leading-order coefficients $(p_{\alpha}^{(k)})_m$ of the minor eigenvalue branch are more difficult to obtain; they are given by the non-zero solutions of
\begin{equation}
    \sum_{(\mu, \nu) \in \mathcal{N}_k} \left(p_{\alpha}^{(k)} \right)^{-\nu} b_{\mu, \nu} = 0 
    \label{eq:p_alpha_i}
\end{equation}
where $\mathcal{N}_k$ is the set of carrier points lying on the $k$-th relevant line segment, and where $b_{\mu, \nu}$ are the coefficients of the characteristic polynomial in Eq.~\eqref{eq:cPs}.
Note that $b_{0,0} = 1$.

Note also that, for the \mbox{$k$-th} relevant line segment with leading-order fractional power $\kappa_{\alpha}^{(k)}$---and thus, for the \mbox{$k$-th} main eigenvalue branch---there exist, in general, multiple minor eigenvalue branches due to the multiple solutions $(p_{\alpha}^{(k)})_m$ of Eq.~\eqref{eq:p_alpha_i}.
For instance, applying Eq.~\eqref{eq:p_alpha_i} to the carrier shown in Fig.~\ref{fig:carrierAndPolygon}(e) will lead to the quadratic equation $(p_{\alpha}^{(1)})^2 + b_{1,1} (p_{\alpha}^{(1)}) + b_{2,2} = 0$, since $b_{0,0} = 1$. The quadratic equation has two solutions, namely $(p_{\alpha}^{(1)})_1$ and $(p_{\alpha}^{(1)})_2$, for the first and only relevant line segment. 

To better understand the above concepts, let us focus on the instructive and most simple case where there are only two carrier points lying on the $k$-th relevant line segment, which thus must be its starting point $(\mu_{k-1},\nu_{k-1})$ and its end point $(\mu_k,\nu_k)$.
This scenario gives simple solutions to \cref{eq:p_alpha_i}, which also nicely illustrate the arising of different minor eigenvalue branches; the solutions read
\begin{equation}
    (p_{\alpha}^{(k)})_m = \left( - \frac{b_{\mu_k, \nu_k}}{b_{\mu_{k-1}, \nu_{k-1}}} \right)^{\frac{1}{\nu_k - \nu_{k-1}}} e^{2 \pi i \frac{m}{\nu_k - \nu_{k-1}}} , \label{eq:p_alpha_i_special}
\end{equation}
where $m = 1, ..., \nu_k - \nu_{k-1}$. 
Since $b_{0,0} = 1$, \cref{eq:p_alpha_i_special} becomes for the first relevant line segment ($k=1$)
\begin{equation}
(p_{\alpha}^{(1)})_m = (- b_{\mu_1, \nu_1} )^{\frac{1}{\nu_1}} e^{2 \pi i \frac{m}{\nu_1}} . \label{eq:p_alpha_i_special_first}
\end{equation}
Let us now characterize the type of perturbation in a basis-independent way.
If we consider a Hamiltonian in the form of Eq.~\eqref{eq:effective_hamiltonian} with dimension $n$ where the unperturbed part $H_0$ exhibits an EP of order $n$, the condition for a generic perturbation (that is, eigenvalues scale proportional to $\epsilon^{\frac{1}{n}}$) can be specified with Eq.~\eqref{eq:kappa_alpha_i} in terms of the presence of the coefficient of the characteristic polynomial
\begin{equation}
    b_{1, n} \neq 0 . \label{eq:generic_condition}
\end{equation}
If this coefficient equals zero, a non-generic perturbation is present.
The condition in Eq.~\eqref{eq:generic_condition}, in combination with the result $(p_{\alpha}^{(1)})_m = (- b_{1, n} )^{\frac{1}{n}} e^{2 \pi i \frac{m}{n}}$ with $m=1,2, ..., n$ of Eq.~\eqref{eq:p_alpha_i_special_first} as well as the structure of the Puiseux series in Eq.~\eqref{eq:puiseux_struc} can be seen as the basis-independent version of Theorem 1 in Ref.~\cite{2012_demange}. 
This comes at the prize of calculating the coefficient $b_{1, n}$ of the characteristic polynomial instead of checking one element of the perturbation matrix in the Jordan-normal basis. 
However, as we will show in \cref{sec:graph_quantities}, the calculation of the coefficients $b_{\mu, \nu}$ can conveniently be done by considering a graph representation of the effective Hamiltonian $H$.

\subsection{Searching for the relevant coefficients of the characteristic polynomial} \label{sec:searching_seq}
So far, we mainly specified the results concerning the Puiseux series of Ref.~\cite{1950_walker} for the characteristic polynomial, \cref{eq:cPs}, arising from the Hamiltonian described by Eq.~\eqref{eq:effective_hamiltonian}.
An important insight, which can already be recognized in \cref{fig:carrierAndPolygon}, motivates our graph-theoretical approach. 
Typically, it is not necessary to have information about all carrier points to construct the relevant line segments of the Newton polygon. 
If one has the possibility to calculate the coefficients of the characteristic polynomial $b_{\mu, \nu}$ independently, a searching sequence can be carried out which delivers just enough information to determine the leading-order terms of the Puiseux series without the knowledge of all $b_{\mu, \nu}$. 
In the following \cref{sec:graph_quantities}, we present a method based on graph theory that achieves this. The derivation of the method is shown in \cref{app:char_poly}.

Before proceeding to the next section, we will discuss the searching sequence mentioned above.
The idea is to prioritize the order in which the $b_{\mu, \nu}$ are checked for their equivalence to zero.
We remind the reader that the non-vanishing coefficients form the carrier of the characteristic polynomial of the underlying Hamiltonian $H$, and that each coefficient $b_{\mu,\nu}$ corresponds to the carrier point $(\mu,\nu)$.
Roughly speaking, having all potential carrier points in mind, we start checking the $b_{\mu, \nu}$ where the slope of the corresponding line segment would be the steepest positive one; this follows from the definition of the relevant line segments given in \cref{sec:newton_polygon}.
If a $b_{\mu, \nu}$ equals zero, then the corresponding point $(\mu,\nu)$ is not part of the carrier $\Delta P_{\text{char}}$ and hence not part of a relevant line segment.
On the other hand, if $b_{\mu, \nu}$ is unequal to zero, we directly obtain a part of the necessary information to calculate the leading-order term with \cref{eq:kappa_alpha_i,eq:p_alpha_i}. 

Concretely, let us consider a general $4\times4$ Hamiltonian with the structure from Eq.~\eqref{eq:effective_hamiltonian}.
We remind the reader that, as a result of \cref{eq:effective_hamiltonian}, all potential carrier points arise in the triangular shape as shown in \cref{fig:carrierAndPolygon}(a).
Using the above mentioned criterion of the steepest slope, we depict the ordering for the first relevant line segment in \cref{fig:carrierAndPolygon}(f).
Specifically, the figure shows six lines; their number corresponds to the evaluation order.
If a given grid point $(\mu,\nu)$ lies on the $n$-th line, then one has to evaluate its corresponding coefficient $b_{\mu,\nu}$ only after all the coefficients that lie on the lines $1,\ldots{},n-1$ have been evaluated to zero.
Thus, we see that, in order to determine the carrier points related to the first relevant line segment $\mathcal{S}_1$, we check first the coefficient $b_{1, 4}$ because the point $(1,4)$ lies on the first line.
The next step depends on the exact values of the matrix elements of \cref{eq:effective_hamiltonian}.
If $b_{1,4}$ is unequal to zero [as in \cref{fig:carrierAndPolygon}(c)], we have found the first relevant line segment $\mathcal{S}_1$ and can determine the leading-order behavior with \cref{eq:kappa_alpha_i,eq:p_alpha_i}. 
If, on the other hand, $b_{1,4}$ equals zero [as in \cref{fig:carrierAndPolygon}(d)], we need to continue our searching sequence.
The next coefficient to check is then $b_{1,3}$, since this point is the only one lying on the second line in \cref{fig:carrierAndPolygon}(f).
Again, if $b_{1,3}$ is non-vanishing, we have found the first relevant line segment $\mathcal{S}_1$; this is the case for \cref{fig:carrierAndPolygon}(d).

To continue with the (potentially existing) second relevant line segment $\mathcal{S}_2$, we would initiate the searching sequence anew, with the end point of the first relevant line segment as starting point.
Staying in the example of \cref{fig:carrierAndPolygon}(d), this would be the point $(1,3)$.
Again using the condition of the steepest slope, the next coefficient to be checked would be $b_{2,4}$; the slope of the potential line segment between $(1,3)$ and $(2,4)$ would be equal to unity.
In the case of \cref{fig:carrierAndPolygon}(d), $b_{2,4}$ is indeed unequal to zero and we have found the second relevant line segment $\mathcal{S}_2$.
Note that, at this point, we are guaranteed to have found all relevant line segments, since the next possible carrier points would be $(3,4)$ and $(3,3)$; the resulting line segments would have vanishing or negative slopes, respectively, and are thus not relevant.
Let us now go back a bit to the point where we evaluate $b_{1,3}$.
If, unlike in \cref{fig:carrierAndPolygon}(d), this coefficient does vanish [as in \cref{fig:carrierAndPolygon}(e)], we need to continue the search for the first relevant line segment $\mathcal{S}_1$.
Thus, we next have to check the coefficients corresponding to the grid points lying on lines $3$, $4$, $5$, and $6$.
Looking at \cref{fig:carrierAndPolygon}(f), we see that the order for which the coefficients are checked is given by $(b_{1,2}, b_{2,4})$ with corresponding slope of 2, next $b_{2,3}$ with slope of $3/2$, then $b_{3,4}$ with slope $4/3$, and finally $(b_{1,1}, b_{2,2}, b_{3,3}, b_{4,4})$ with slope $1$.
In the case of \cref{fig:carrierAndPolygon}(e), only the coefficients $b_{1,1}$ and $b_{2,2}$ are unequal to zero.
Note that there is no possibility of a second relevant line segment $\mathcal{S}_2$, since we have checked all possible grid points that the structure of \cref{eq:effective_hamiltonian} allows to be non-vanishing; in other words, we have checked the whole triangle depicted in \cref{fig:carrierAndPolygon}(a).

There is an imagination of such a sequence for the $k$-th relevant line segment.
One can think of a clock hand where the rotation center is at the end carrier point of the $(k-1)$-th relevant line segment $(\mu_{k-1}, \nu_{k-1})$ in the $\mu$-$\nu$-plane. The starting time of the algorithm is 12'o clock; in other words, the clock hand shows in positive $\nu$-direction.
As time progresses, we will eventually hit one or more potential carrier points of the characteristic polynomial. This moment freezes the time and therefore the clock hand position.
Now all $b_{\mu, \nu}$ corresponding to the points $(\mu, \nu)$ that lie on the clock hand (with the exception of the clock hand center) have to be evaluated.
If at least one of these $b_{\mu, \nu}$ is unequal to zero, the searching sequence for the $k$-th relevant line segment is finished. Otherwise, the time continues to run until the clock hand hits the next carrier point(s) and the process repeats. 
Later on, in \cref{sec:leading_puiseux}, we depict this ``running of the clock'' visualization of the searching sequence. 

By performing the searching sequence, one obtains all information needed to determine the leading-order terms of the Puiseux series with \cref{eq:kappa_alpha_i,eq:p_alpha_i}.

We remark that, during the searching sequence, the first evaluation is performed on $b_{1, n}$. For a perturbed EP of order $n$ in an $n\times n$ Hamiltonian, this is equivalent to checking the condition in Eq.~\eqref{eq:generic_condition} for the presence of a generic perturbation. Hence, after the first evaluation, a classification of the perturbation in generic or non-generic is obtained. 
The following evaluations (if necessary) refine the classification in terms of the asymptotically leading-order behavior of the eigenvalue branches. 

\subsection{Finding the coefficients of the characteristic polynomial with LSDs} \label{sec:graph_quantities}
As we have seen in~\cref{sec:newton_polygon,sec:searching_seq}, the leading-order term of the Puiseux series in Eq.~\eqref{eq:branches-k-segment} is determined by the relevant line segments of the Newton polygon, which in turn can be constructed from the coefficients of the characteristic polynomial.
Let us now link these coefficients to graph theory. 

We consider the characteristic polynomial calculated with the combinatorial formula \cite{1990Marcus,Brualdi2008CombinatorialApproachMatrixTheory}
\begin{equation}
    \det (\lambda \mathbb{1} - H) = \sum_{\nu = 0}^n (-1)^{\nu} c_{\nu} (H) \lambda^{n - \nu} \label{eq:char_poly_combi_main}
\end{equation}
where $c_{\nu}$ equals the sum of all principal minors of order~$\nu$ of the Hamiltonian $H$. A principal minor of order $\nu$ is defined as the determinant of a submatrix of~$H$ with dimension $\nu$ where $n- \nu$ rows and columns with the same indices are deleted appropriately. These submatrices can be mapped one-to-one into the concept of induced subgraphs that contain only $\nu$ out of the $n$ vertices of the original graph; see \cref{sec:graph_foundation}.
Moreover, with the help of the Coates determinant formula in \cref{eq:coates_det_main} each summand of the principal minor $c_{\nu}$ can be written as a sum over the LSDs.
This is the main idea for deriving (see \cref{app:char_poly} for details)
 \begin{equation}
    b_{\mu, \nu} = \sum_{L \in \mathcal{L}_{\mu, \nu} (H)} (-1)^{c(L)} w(L) \,.
    \label{eq:b_mu_nu_graph_main}
\end{equation}
Here, $\mathcal{L}_{\mu, \nu} (H)$ describes the set of all LSDs from all induced subgraphs of $H$ consisting of $\nu$ vertices where $\mu$-times an $\epsilon$-edge is involved. This concept is explained in the next paragraph.
The number of disconnected cycles of a LSD $L$ is $c(L)$ and the product of the edge weights of a LSD is $w(L)$. 

\begin{figure}[tb]
	\begin{center}
		\includegraphics[width=1.0\columnwidth]{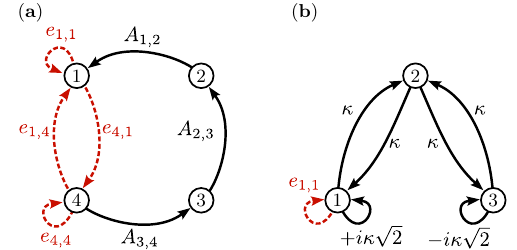}
		\caption{Examples of multi digraphs used in \cref{eq:b_mu_nu_graph_main} for (a) the example Hamiltonian in \cref{eq:H_example_2ring} and 
			(b) the $\mathcal{PT}$-symmetric system studied in Ref.~\cite{HHW17}. 
		}
		\label{fig:modified_coates_digraph}
	\end{center}
\end{figure}
\begin{figure}[tb]
	\begin{center}
		\includegraphics[width=1.0\columnwidth]{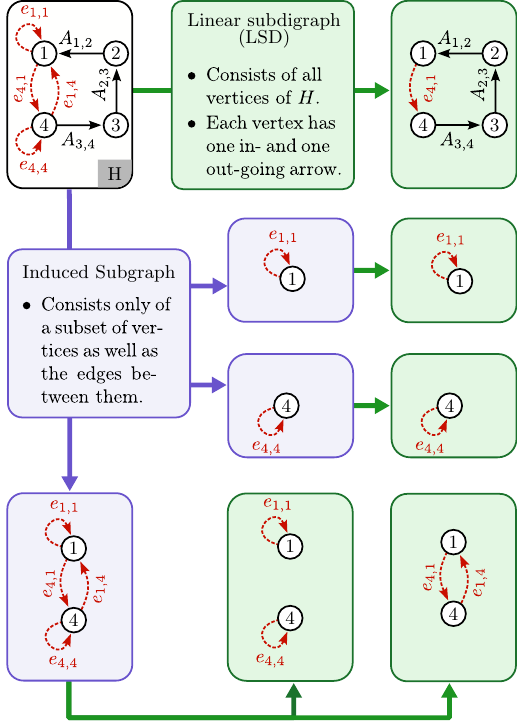}
		\caption{Visualization of the concepts of LSDs and induced subgraphs by means of the graph corresponding to the matrix~$H$ of \cref{eq:H_example_2ring}. Not all possible induced subgraphs are shown, only those that possess at least one LSD. The induced subgraphs consisting of a single vertex are here equal to their respective LSDs. The induced subgraph consisting of two vertices possesses two LSDs. 
		} 
		\label{fig:scheme_graph_quantities_example}
	\end{center}
\end{figure}

The combinatorial complexity of \cref{eq:b_mu_nu_graph_main} originates from the set $\mathcal{L}_{\mu, \nu} (H)$. 
In order to be able to understand this set, we define a directed graph for which it is possible to have two edges starting and ending from the same vertex. It is a so-called multi graph, which can be seen as a modified Coates digraph used in the Coates determinant formula in~\cref{eq:coates_det_main}. 
An example of such a multi graph is depicted in \cref{fig:modified_coates_digraph} (a).
It shows the example Hamiltonian of \cref{eq:H_example_2ring} such that solid black edges correspond to matrix elements of $H_0$, while dashed red edges correspond to matrix elements of $H_1$.
Note that in the given definition, the perturbation parameter $\epsilon$ does not appear in the edge weights. 
Another multi digraph is depicted in \cref{fig:modified_coates_digraph} (b), showcasing the possibility of two edges from a vertex to itself.
Overall, the concept of a multi digraph allows us to distinguish, on the level of the graph, between edges which are related to the unperturbed part $H_0$, and edges that are related to the perturbation matrix $H_1$.
In the following, we use the terminology $\epsilon$-edge for the latter, because they arise due to the perturbation.
\begin{figure*}[tb]
	\begin{center}
		\includegraphics[width=1.0\textwidth]{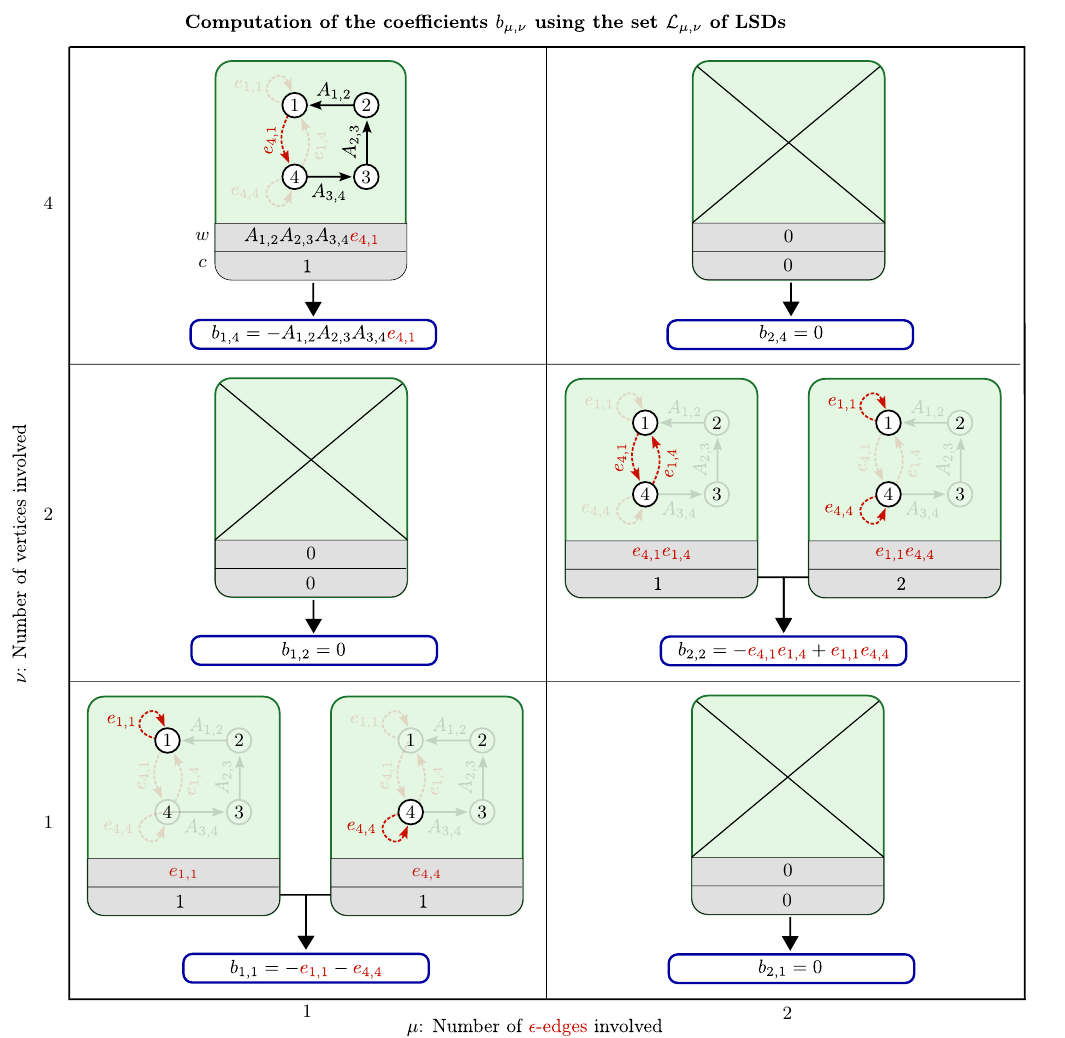}
		\caption{
  The set of LSDs $\mathcal{L}_{\mu, \nu} (H)$ of the example Hamiltonian in \cref{eq:H_example_2ring} 
  is visualized. The graph representation of the Hamiltonian (also shown in \cref{fig:scheme_graph_quantities_example}) is indicated with all vertices and edges (toned down and highlighted). Each cell of the table contains all the corresponding LSDs (highlighted) of the set $\mathcal{L}_{\mu, \nu} (H)$; crossed boxes mean that there are no LSDs in the set. Each LSD has two properties: The product~$w$ of its weights, and the number $c$ of components (these are set to zero if there are no LSDs). These are used to obtain the coefficients of the characteristic polynomial $b_{\mu, \nu}$ via \cref{eq:b_mu_nu_graph_main}; the result is written in the blue boxes at the bottom of each cell of the table. The row $\nu = 3$ is omitted since both $\mathcal{L}_{1, 3} (H)$ and $\mathcal{L}_{2, 3} (H)$ are empty sets; as a result, $b_{1,3} = b_{2,3} = 0$. The columns with $\mu = 3,4$ are omitted for the same reason.
  } 
		\label{fig:scheme_b_mu_nu_example}
	\end{center}
\end{figure*}
By a combination of the two concepts of induced subgraphs and LSDs, we obtain the elements of the set $\mathcal{L}_{\mu, \nu} (H)$. 
In a first step, one draws all possible induced subgraphs of $H$ that contain exactly $\nu$ vertices.
In a second step, all possible LSDs of these induced subgraphs are drawn; the ones that contain $\mu$-times an $\epsilon$-edge form the set $\mathcal{L}_{\mu, \nu} (H)$.
To get acquainted with this logic, in \cref{fig:scheme_graph_quantities_example}, we showcase all possible induced subgraphs that have at least one LSD and also draw these LSDs for the example Hamiltonian $H$ given by \cref{eq:H_example_2ring} in \cref{sec:puiseux_foundation}.

After understanding the set $\mathcal{L}_{\mu, \nu} (H)$ better, we can now apply \cref{eq:b_mu_nu_graph_main} to the example mentioned above.
In graphical form, this application is depicted in \cref{fig:scheme_b_mu_nu_example}.
In this figure, all LSDs obtained in \cref{fig:scheme_graph_quantities_example} are grouped by the number of vertices they cover, $\nu,$ and the number of $\epsilon$-edges involved, $\mu$.

With \cref{eq:b_mu_nu_graph_main} we set the backbone of our graph-theoretical approach. By finding LSDs as described above, it empowers us to determine the coefficients of the characteristic polynomial $b_{\mu, \nu}$ independently. 
We want to highlight that Eq.~\eqref{eq:b_mu_nu_graph_main} is valid for any matrix with the structure $A = B + \epsilon\, C$.
In \cref{app:multi_par} is discussed how the coefficients of the characteristic polynomial in multiple variables, namely the eigenvalue and various system parameters, can be determined independently in a similar fashion. 

\subsection{The unified graph picture} \label{sec:leading_puiseux}

\begin{figure*}[tb]
	\begin{center}
		\includegraphics[width=1.0\textwidth]{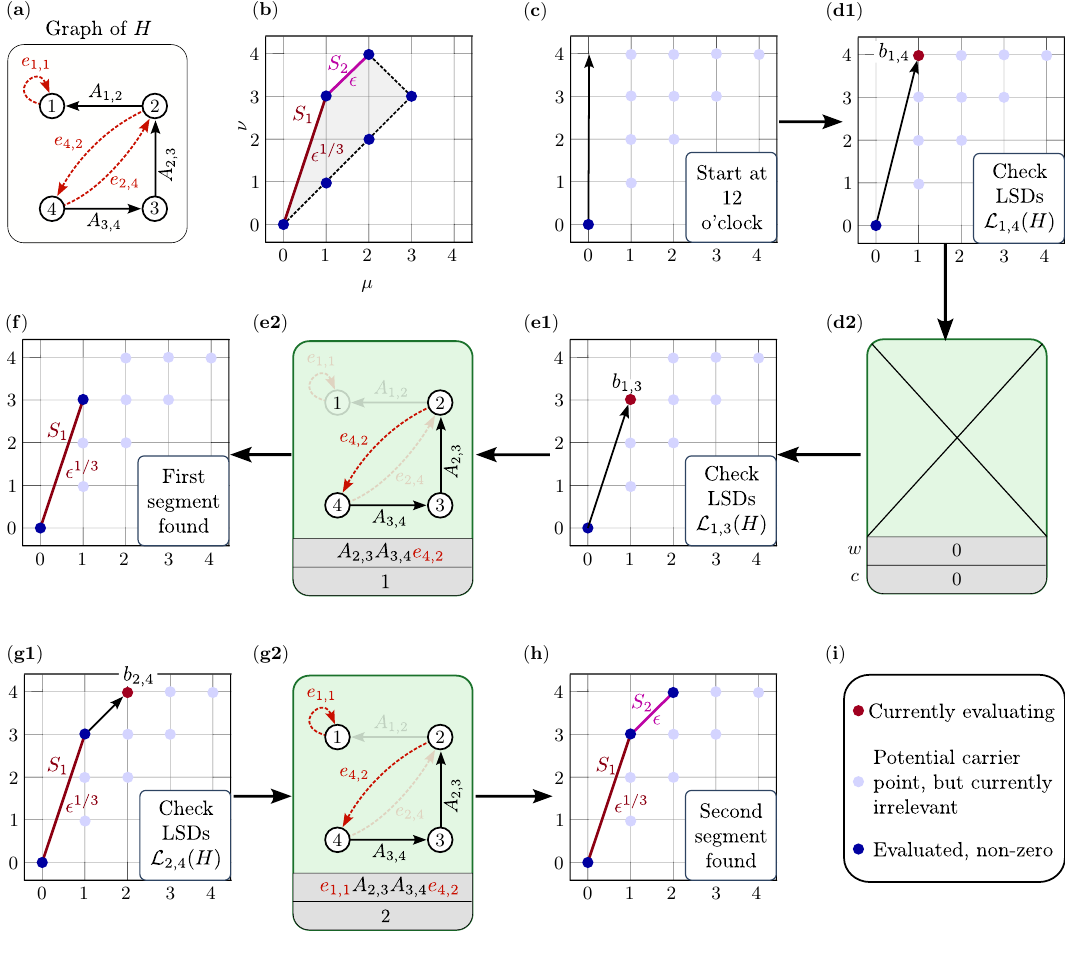}
		\caption{
  Exemplary use of the searching sequence described in \cref{sec:searching_seq} combined with graph theory described in \cref{sec:graph_quantities}. (a) Graph representation of an example Hamiltonian. (b) Corresponding carrier with the Newton polygon and the relevant line segments. (c)-(h) Visualization of the searching sequence. 
  (c) Initialisation of the searching sequence. (d1), (e1), and (g1) Visualization of the potential carrier points which have to be evaluated. (d2), (e2), and (g2) Evaluation with graph theory [Eq.~\eqref{eq:b_mu_nu_graph_main} schematically explained in \cref{fig:scheme_b_mu_nu_example}]. (f) and (h) The relevant line segments found with the searching sequence. 
  (i) Legend for the carrier points in (b), (c), (d1), (e1), (f), (g1), and (h).} 
    \label{fig:lin_sub_digraph_2ring_non_generic_artificial}
	\end{center}
\end{figure*}

In previous sections, we introduced a variety of concepts to determine the leading-order term of the Puiseux series (\cref{sec:newton_polygon}) under consideration of the relevant coefficients of the characteristic polynomial (\cref{sec:searching_seq}), as well as the graph-theoretical formula in \cref{eq:b_mu_nu_graph_main} to obtain the coefficients of the characteristic polynomial independently (\cref{sec:graph_quantities}).
Here, these concepts are combined to provide a unified graph-theoretical picture. 

Given a perturbed Hamiltonian $H(\epsilon)$ as described in \cref{eq:effective_hamiltonian}, our graph-theoretical approach gives---independent of the basis chosen for $H$---the eigenvalue behavior for small perturbations. Additionally, it allows to classify the perturbative behavior into generic or non-generic. 
Explicitly, the approach consists of the following steps:
\begin{enumerate}
    \item Determine the set of potentially relevant index-tuples~$(\mu, \nu)$ with the steepest slope condition as described in \cref{sec:searching_seq}.
    \item For each index tuple $(\mu, \nu)$, apply \cref{eq:b_mu_nu_graph_main} by finding LSDs with $\nu$ vertices covered while $\mu$-times an $\epsilon$-edge is involved to determine $b_{\mu, \nu}$.
    \item Determine the number of relevant index-tuples~$(\mu, \nu)$ with non-vanishing $b_{\mu, \nu}$. If there are exactly two such tuples, use \cref{eq:kappa_alpha_i,eq:p_alpha_i_special} to determine the leading-order terms $(p_{\alpha}^{(k)})_m \epsilon^{\kappa_{\alpha}^{(k)}}$ of the Puiseux series in \cref{eq:branches-k-segment}. If there are more than two tuples, use \cref{eq:kappa_alpha_i,eq:p_alpha_i}. Otherwise, start again from the first step by searching for the next set of potentially relevant index-tuples~$(\mu, \nu)$.
\end{enumerate}

We consider two examples to illustrate the usage and insights of the approach. 
Let us first have a look at the example which accompanied us since \cref{sec:puiseux_foundation}; that is, the one described by the Hamiltonian $H$ of \cref{eq:H_example_2ring}. 
We find the first potentially relevant tuple to be $(1, 4)$ because this leads to the maximum positive slope of 4 of the line segments starting from $(0,0)$ Analogously, one could also search for the LSD with the most vertices covered while the least $\epsilon$-edges involved. 
As we have seen in \cref{fig:scheme_b_mu_nu_example}, by finding the LSD which is the only element of the set $\mathcal{L}_{1, 4} (H)$ we obtain $b_{1, 4} = -A_{1,2} A_{2,3} A_{3,4} e_{4,1}$ due to \cref{eq:b_mu_nu_graph_main}. This leads us to the usage of \cref{eq:kappa_alpha_i,eq:p_alpha_i_special_first} which results in the eigenvalue branches in leading order given by \cref{eq:puiseux_example_generic}. Note that $b_{0, 0} = 1$.

Next, we consider a more elaborate example which leads to the carrier shown in \cref{fig:carrierAndPolygon}(d) [again depicted in \cref{fig:lin_sub_digraph_2ring_non_generic_artificial}(b) for convenience], where two different leading-order fractional powers $\kappa_{\alpha}^{(1,2)}$---and thus two main eigenvalue branches---are present.
A Hamiltonian in graph representation which leads to such a carrier is shown in \cref{fig:lin_sub_digraph_2ring_non_generic_artificial}(a). Here, $H_0$ is equal to the previous example; only the perturbation~$H_1$ has changed. 
Since four vertices are present, we first search for LSDs of the set $\mathcal{L}_{1,4} (H)$ as visualized in \cref{fig:lin_sub_digraph_2ring_non_generic_artificial}(c) and~(d1). No such LSD is present, see \cref{fig:lin_sub_digraph_2ring_non_generic_artificial}(d2), and hence we continue to search for the LSDs in the set $\mathcal{L}_{1,3} (H)$ due to the steepest slope condition as shown in \cref{fig:lin_sub_digraph_2ring_non_generic_artificial}(e1). 
At this point, we already know that a non-generic perturbation is present because $\kappa_{\alpha}$ is unequal to $1/4$.
We find exactly one LSD in the set $\mathcal{L}_{1,3}(H)$, see \cref{fig:lin_sub_digraph_2ring_non_generic_artificial}(e2), which gives, with \cref{eq:b_mu_nu_graph_main}, $b_{1,3} = - A_{2,3} A_{3,4} e_{4,2}$. Via \cref{eq:kappa_alpha_i,eq:p_alpha_i_special_first}, we find the three minor eigenvalue branches arising from the main branch related to the first relevant line segment [\cref{fig:lin_sub_digraph_2ring_non_generic_artificial}(f)] to be $\lambda^{(1)}_m = (A_{2,3} A_{3,4} e_{4,2} e^{2 \pi i m} \epsilon)^{\frac{1}{3}}$, where $m = 1,2,3$.
For the next searching step, we set $(\mu_1, \nu_1) = (1, 3)$ and through the steepest slope condition we know that we have to search for LSD in $\mathcal{L}_{2,4}(H)$ which is emphasized in \cref{fig:lin_sub_digraph_2ring_non_generic_artificial}(g1). Exactly one LSD of the set is present [\cref{fig:lin_sub_digraph_2ring_non_generic_artificial}(g2)], which leads to $b_{2,4} = A_{2,3} A_{3,4} e_{4,2} e_{1,1}$. We obtain the remaining eigenvalue branch related to the second relevant line segment [\cref{fig:lin_sub_digraph_2ring_non_generic_artificial}(h)] with \cref{eq:kappa_alpha_i,eq:p_alpha_i_special} to be $\lambda^{(2)}_1 = e_{4,4} \epsilon$. Here, the main branch has no subdivision into minor branches due to the single solution of Eq.~\eqref{eq:p_alpha_i_special}. 

We state one important insight of the graph-theoretical approach concerning higher-order EPs, explicitly. 
Roughly speaking, the eigenvalue behavior near EPs---which corresponds to the leading-order term in the Puiseux series---is determined by the LSDs where the most vertices of the graph representation of the Hamiltonian are covered while the least $\epsilon$-edges are involved. 

\section{Revisiting coupled optical microring systems exhibiting higher-order EPs} \label{sec:physical_examples}
To fill the graph-theoretical approach with physical meaning, we apply it to a particular interesting class of systems, namely coupled optical microrings. These systems exhibit higher-order EPs which have
been experimentally realized for sensing applications~\cite{HHW17} and theoretically proposed with a special attention to the robustness considering fabrication errors~\cite{2023_kullig}.

\subsection{$\mathcal{PT}$-symmetric system with three microrings: Heating perturbation} \label{sec:pt-sym}

In this section, the $\mathcal{PT}$-symmetric system from Ref.~\cite{HHW17} consisting of three evanescently coupled microrings is considered. This system is perturbed by slightly changing the eigenfrequency of one microring by heating the microring. The applied perturbation leads to a generic behavior of the third-order EP. 

\begin{figure}[tb]
	\begin{center}
		\includegraphics[width=1.0\columnwidth]{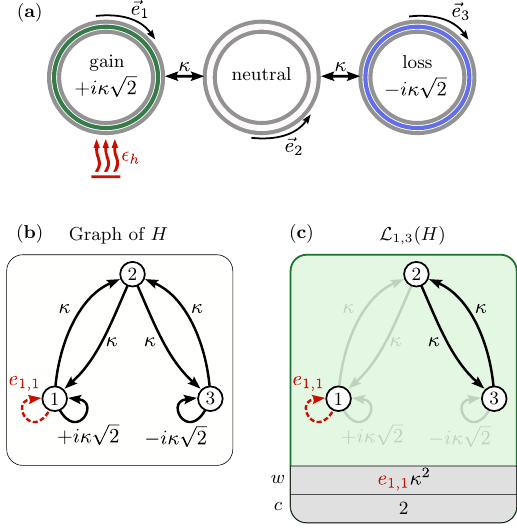}
\caption{(a) Schematic view of the $\mathcal{PT}$-symmetric system from Ref.~\cite{HHW17} with Hamiltonian~$H$ in Eq.~(\ref{eq:HPT}). The parameter for the bidirectional coupling between microrings is given by~$\kappa$. The microring intrinsic gain (green) and loss (blue) is given by $\pm i \sqrt{2} \kappa$, respectively. The heating-induced perturbation of the left microring (red) is modelled with $\epsilon_he_{1,1}$. The basis of~$H$ is chosen as the traveling waves indicated by the arrows $\vec{e}_{i}$ with $i = 1,2,3$. (b) The graph representation of~$H$. Black edges correspond to the unperturbed part of~$H$ and red edges correspond to the perturbation. (c) The only LSD (highlighted) of the set $\mathcal{L}_{1,3} (H)$, which comprises the LSDs with one $\epsilon$-edge involved and three vertices covered. Toned down and highlighted edges form the graph representation of the Hamiltonian as shown in (b).
}
		\label{fig:lin_sub_digraphs_enhanced_sens}
	\end{center}
\end{figure}

The graph representation of the effective Hamiltonian
\begin{equation}\label{eq:HPT}
    H = \left( \begin{array}{c c c}
        i \sqrt{2} \kappa & \kappa & 0 \\
        \kappa & 0 & \kappa \\
        0 & \kappa & - i \sqrt{2} \kappa 
    \end{array} \right)
    + \epsilon_{h}
    \left( \begin{array}{c c c}
        e_{1,1} & 0 & 0 \\
        0 & 0 & 0 \\
        0 & 0 & 0 
    \end{array} \right)
\end{equation}
of the $\mathcal{PT}$-symmetric system is shown in \cref{fig:lin_sub_digraphs_enhanced_sens}(b).
A schematic representation of the physical system and its relation to the effective Hamiltonian is given in Fig.~\ref{fig:lin_sub_digraphs_enhanced_sens}(a). Here, the canonical basis for the effective Hamiltonian $\vec{e}_i$ with $i = 1, 2, 3$ is chosen as the traveling wave basis of the single microrings, where waves travel in clockwise (CW) or counterclockwise (CCW) direction in the cavity. In the graph picture in Fig.~\ref{fig:lin_sub_digraphs_enhanced_sens}(b), the basis $\vec{e}_i$ corresponds to the vertex labeled by $i$. The evanescent coupling between the microrings is described by the parameter $\kappa \in \mathbb{R}$. The gain and loss of the cavities at the third-order EP is given by the parameter $\pm i \sqrt{2} \kappa$. The couplings are the black edges in the graph picture where their weight is given by the coupling parameter. 
The coupling $\epsilon_h$ induced by the perturbation corresponds to the red edge in the graph representation. 

The natural choice of the traveling wave basis in combination with the couplings enables one to link directly the schematic representation of the system [Fig.~\ref{fig:lin_sub_digraphs_enhanced_sens}(a)] to its graph representation [Fig.~\ref{fig:lin_sub_digraphs_enhanced_sens}(b)] which is a one-to-one map of the effective Hamiltonian in Eq.~\eqref{eq:HPT}.
Furthermore, the graph picture can be intuitively interpreted in a physical manner.  
It can be seen as a map which shows all possible paths of the light in the system. With this interpretation, the graph theoretical approach filters specific light paths which are responsible for the perturbative behavior near EPs. 

Starting from the graph representation of the Hamiltonian in \cref{fig:lin_sub_digraphs_enhanced_sens}(b), our approach explained in Section~\ref{sec:leading_puiseux} is performed, with $(1,3)$ being the first grid point to be checked.
The highlighted LSD in Fig.~\ref{fig:lin_sub_digraphs_enhanced_sens}(c) is the only element of the set $\mathcal{L}_{1,3} (H)$, which, we remind the reader, comprises of LSDs with one $\epsilon$-edge involved and three vertices covered.
This single LSD can be interpreted as a light wave which travels between the second and third cavity and a light wave which travels only in the first cavity. 
The parameters of this graph are $w = e_{1,1} \kappa^2$, and $c=2$. Thus, using Eq.~\eqref{eq:b_mu_nu_graph_main}, we obtain $b_{1, 3} =  e_{1,1} \kappa^2$ and via \cref{eq:p_alpha_i_special_first}, the eigenvalue branches in leading order are given by $\lambda_m (\epsilon) = (- \kappa^2 e_{1,1} \epsilon e^{2 \pi i m})^{\frac{1}{3}} + \ldots$ with $m=1,2,3$. The perturbation is generic.

\subsection{Three waveguide-coupled microrings: Heating perturbation} \label{sec:heatingPerturbation}
\begin{figure}[!tb]
	\begin{center}
		\includegraphics[width=1.0\columnwidth]{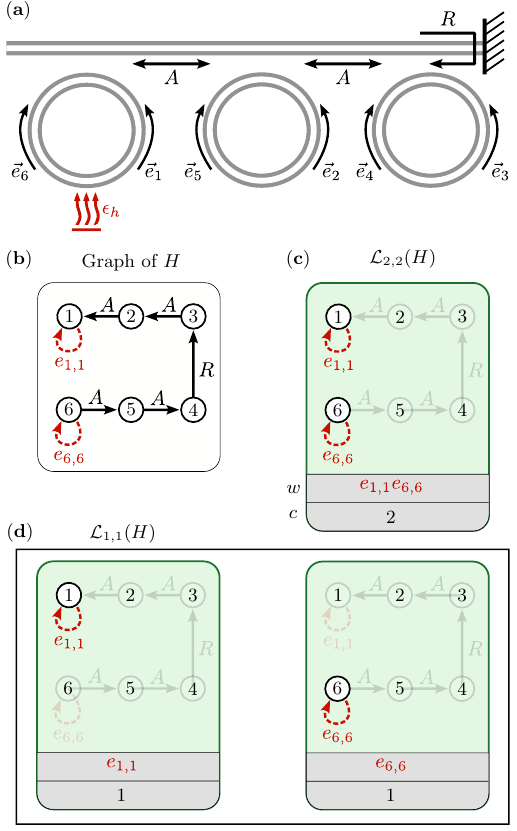}
\caption{(a) Schematic view of the three waveguide-coupled microrings with mirror-induced asymmetric backscattering from Ref.~\cite{2023_kullig} where the parameters of the Hamiltonian~$H$ in Eqs.~(\ref{eq:H0_wg-3ring}) and~(\ref{eq:H1_heating_wg-3ring}) are indicated. The parameter for the coupling between microrings via the waveguide is given by $A$. The mirror-induced coupling is modelled with $R$. The heating-induced perturbation of the left microring (red) is modelled with~$\epsilon_h$. The basis of~$H$ is chosen as the traveling waves and indicated with the arrows~$\vec{e}_{i}$ with $i = 1,2, \ldots,6$. (b) The graph representation of~$H$ of the system. Black edges correspond to the unperturbed part of~$H$ and red edges correspond to the perturbation. (c) shows the only LSD (highlighted) in $\mathcal{L}_{2,2} (H)$, whereas (d) shows all LSDs (highlighted) in the set $\mathcal{L}_{1,1} (H)$. These LSDs are via Eq.~\eqref{eq:b_mu_nu_graph_main} related to the coefficients of the characteristic polynomial. Toned down and highlighted edges form the graph representation of the Hamiltonian as shown in (b)
}
		\label{fig:lin_sub_digraphs_3ring-wg_heating}
	\end{center}
\end{figure}
The system consisting of three waveguide-coupled microrings with mirror-induced asymmetric backscattering~\cite{2023_kullig}
is shown in Fig.~\ref{fig:lin_sub_digraphs_3ring-wg_heating}(a). 
The unperturbed Hamiltonian of the system is given by
\begin{equation}
    H_0^{\text{EP}_6} = \left( \begin{array}{c c c c c c}
        0 & A & 0 & 0 & 0 & 0 \\
        0 & 0 & A & 0 & 0 & 0 \\
        0 & 0 & 0 & R & 0 & 0 \\
        0 & 0 & 0 & 0 & A & 0 \\
        0 & 0 & 0 & 0 & 0 & A \\
        0 & 0 & 0 & 0 & 0 & 0 
    \end{array} \right) \label{eq:H0_wg-3ring}
\end{equation}
and exhibits an EP of order 6 which becomes apparent from the quasi-Jordan-normal form of \cref{eq:H0_wg-3ring}.
The traveling-wave basis $\vec{e}_i$ with $i=1, \ldots , 6$ is chosen as the canonical basis of the effective Hamiltonian~$H$. Here, the CW and CCW direction in each microring has to be taken into account. The waveguide-coupling is modelled with the parameter $A \in \mathbb{C}$, and the parameter of the mirror-induced coupling is $R \in \mathbb{C}$. The perturbation effects the two traveling waves in one microring and can be described by the perturbation matrix
\begin{equation}
    H_1^{h} = \left( \begin{array}{c c c c c c}
        e_{1,1} & 0 & 0 & 0 & 0 & 0 \\
        0 & 0 & 0 & 0 & 0 & 0 \\
        0 & 0 & 0 & 0 & 0 & 0 \\
        0 & 0 & 0 & 0 & 0 & 0 \\
        0 & 0 & 0 & 0 & 0 & 0 \\
        0 & 0 & 0 & 0 & 0 & e_{6,6} 
    \end{array} \right) . \label{eq:H1_heating_wg-3ring}
\end{equation}

In \cref{fig:lin_sub_digraphs_3ring-wg_heating}(b), we show the graph representation of the system $H = H_0^{\text{EP}_6} + \epsilon_h H_1^{h}$. Again, we can interpret the graph picture as the possible light paths in the system. 

We note that this graph has only two $\epsilon$-edges, and thus all coefficients $b_{\mu,\nu}$ with $\mu>2$ vanish.
The only possibly non-vanishing coefficients are thus $b_{1,2}$, $b_{1,1}$ and $b_{2,2}$.
The first of these three vanishes, since there is no possible LSD comprising of one $\epsilon$-edge and two vertices.
Thus, the only remaining coefficients are $b_{1,1}$ and $b_{2,2}$.
The relevant sets of LSDs $\mathcal{L}_{1,1}$ and $\mathcal{L}_{2,2}$ are shown in Fig.~\ref{fig:lin_sub_digraphs_3ring-wg_heating}(d) and (c), respectively. 
This leads, with Eq.~\eqref{eq:b_mu_nu_graph_main}, to $b_{1,1} = - e_{1,1} - e_{6,6}$ and $b_{2,2} = e_{1,1} e_{6,6}$. 
A simultaneous vanishing of the coefficients $b_{1,1}$ and $b_{2,2}$ is only possible for $e_{1,1} = e_{2,2} = 0$; this would correspond to the unperturbed case. Both of those coefficients are part of the first line segment with a slope of one. 
Hence, it is clear from \cref{sec:newton_polygon} \cref{eq:kappa_alpha_i} that the leading-order term of the Puiseux series is linear in $\epsilon$. This implies that a non-generic perturbation is present. 
\Cref{eq:p_alpha_i}, which here reads $(p_{\alpha})^2 - ( e_{1,1} + e_{6,6} ) (p_{\alpha}) + e_{1,1} e_{6,6} = 0$, determines the dominant Puiseux coefficients $p_{\alpha}^{(1)}$ for the first line segment. 
The two solutions of this equation are $p_{\alpha}^{+} = e_{1,1}$ and $p_{\alpha}^{-} = e_{6,6}$. If $e_{1,1} = e_{6,6}$ then the two solutions are degenerate. The remaining four branches of the eigenvalues are degenerate at the value zero because the set $\mathcal{L}_{\mu, \nu} (H)$ is empty for $\nu > 2$. The eigenvalue branches $\lambda_m$ read as $\{ e_{1,1} \epsilon, e_{6,6} \epsilon, 0, 0, 0, 0 \}$.

From the above two examples, we see that  perturbing the $\mathcal{PT}$-symmetric system in Fig.~\ref{fig:lin_sub_digraphs_enhanced_sens} and the above described system in Fig.~\ref{fig:lin_sub_digraphs_3ring-wg_heating} in the same way results in a completely different eigenvalue behavior near the EP. On the one hand, the generic behavior for the $\mathcal{PT}$-symmetric system is present. Here, three eigenvalue branches will split away from the EP of 3-rd order with a scaling proportional to $\epsilon^{1/3}$. On the other hand, the waveguide-coupled microring system near an EP of order~6 where two eigenvalue branches split with a scaling proportional to $\epsilon$. The remaining eigenvalues stay degenerate at zero. With the graph-theoretical approach, we were able to trace back the different behaviors of the two systems to specific configurations of cyclic paths which the light waves can travel within the systems, namely the set of LSDs shown in Figs.~\ref{fig:lin_sub_digraphs_enhanced_sens}(c) and \ref{fig:lin_sub_digraphs_3ring-wg_heating}(c)-(d). 

\subsection{Three waveguide-coupled microrings: Single-particle perturbation} \label{sec:single_particle_perturbation}

\begin{figure}[!tb]
	\begin{center}
		\includegraphics[width=1.0\columnwidth]{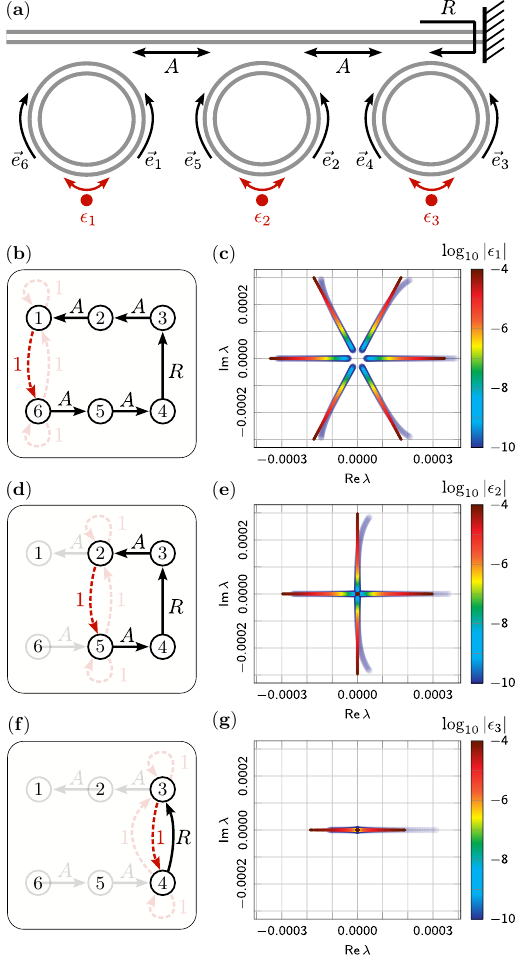}
\caption{(a) Schematic view of the three waveguide-coupled microrings with mirror-induced asymmetric backscattering from Ref.~\cite{2023_kullig} described by the Hamiltonian~$H$ in Eq.~\eqref{eq:Hamiltonian-wg-3ring-single-particle}. 
The basis of~$H$ is chosen as the traveling waves indicated by the arrows~$\vec{e}_{i}$ with $i = 1,2, \ldots,6$. 
(b), (d), and (f) highlight the LSDs which determine the leading order of the Puiseux series for a particle at the left $\epsilon_1$, middle $\epsilon_2$, and right $\epsilon_3$ microring. All edges, highlighted and toned down, show the graph representation of~$H$. (c), (e), and (g) show the dimensionless eigenvalue spectra in the complex plane for the different single-particle perturbations $\epsilon_1$, $\epsilon_2$, and $\epsilon_3$, respectively. The colormap indicates the absolute perturbation strength $|\epsilon_i|$ with $i=1,2,3$ in logarithmic scale. The data shown is obtained by pure real values of $\epsilon_i$, ranging from $10^{-10}$ to $10^{-4}$, $A = 4.7 \cdot 10^{-4}$, and $R = 3.5 \cdot 10^{-4}$.
}
		\label{fig:lin_sub_digraphs_3ring-wg_single-particle}
	\end{center}
\end{figure}

The effect of single-particle perturbations on the waveguide-coupled microring system from Ref.~\cite{2023_kullig} is schematically shown in Fig.~\ref{fig:lin_sub_digraphs_3ring-wg_single-particle}(a).
A single particle in the vicinity of a microring couples the local CW and CCW travelling waves, as indicated by the red double arrows in Fig.~\ref{fig:lin_sub_digraphs_3ring-wg_single-particle}(a). Additionally, the eigenfrequency of the microring is slightly changed. 
Here, only perturbations with one particle near a single microring are considered. 
The corresponding perturbation matrices within the coupled mode theory can be modelled as
\begin{equation}
    H_1^{\text{tp}_1}  =  \left( \begin{array}{c c c c c c}
        1 & 0 & 0 & 0 & 0 & 1 \\
        0 & 0 & 0 & 0 & 0 & 0 \\
        0 & 0 & 0 & 0 & 0 & 0 \\
        0 & 0 & 0 & 0 & 0 & 0 \\
        0 & 0 & 0 & 0 & 0 & 0 \\
        1 & 0 & 0 & 0 & 0 & 1
    \end{array} \right) ,  \label{eq:H1_tp1_wg-3ring} 
\end{equation}
\begin{equation}
    H_1^{\text{tp}_2}  =  \left( \begin{array}{c c c c c c}
        0 & 0 & 0 & 0 & 0 & 0 \\
        0 & 1 & 0 & 0 & 1 & 0 \\
        0 & 0 & 0 & 0 & 0 & 0 \\
        0 & 0 & 0 & 0 & 0 & 0 \\
        0 & 1 & 0 & 0 & 1 & 0 \\
        0 & 0 & 0 & 0 & 0 & 0
    \end{array} \right) ,  \label{eq:H1_tp2_wg-3ring} 
\end{equation}
\begin{equation}
    H_1^{\text{tp}_3}  =  \left( \begin{array}{c c c c c c}
        0 & 0 & 0 & 0 & 0 & 0 \\
        0 & 0 & 0 & 0 & 0 & 0 \\
        0 & 0 & 1 & 1 & 0 & 0 \\
        0 & 0 & 1 & 1 & 0 & 0 \\
        0 & 0 & 0 & 0 & 0 & 0 \\
        0 & 0 & 0 & 0 & 0 & 0
    \end{array} \right)  \label{eq:H1_tp3_wg-3ring} 
\end{equation}
for the single-particle at the left, middle and right microring, respectively.
The effective Hamiltonian of the whole system
\begin{equation}
    H = H_0^{\text{EP}_6} + \epsilon_i H_1^{\text{tp}_i}
    \label{eq:Hamiltonian-wg-3ring-single-particle}
\end{equation}
consists of the unperturbed part $H_0^{\text{EP}_6}$ in \cref{eq:H0_wg-3ring} and one of the three different perturbation matrices [Eqs.~\eqref{eq:H1_tp1_wg-3ring}-\eqref{eq:H1_tp3_wg-3ring}] denoted with the indices $i = 1, 2, 3$.

The eigenvalue spectrum in the complex plane for the single-particle perturbations 
Eqs.~\eqref{eq:H1_tp1_wg-3ring}, \eqref{eq:H1_tp2_wg-3ring}, and \eqref{eq:H1_tp3_wg-3ring}
are shown in Fig.~\ref{fig:lin_sub_digraphs_3ring-wg_single-particle}(c), (e), and (g), respectively. 
The underlying eigenvalue branches visualized as shadows correspond to the numerical eigenvalue solution of the Hamiltonian. 
The colored branches are obtained by the Puiseux series in leading order which is governed by the neighboring LSDs in Fig.~\ref{fig:lin_sub_digraphs_3ring-wg_single-particle}(b), (d), and (f). The edges corresponding to these LSDs are highlighted whereas all visualized edges (toned down and highlighted) correspond to the graph representation of the Hamiltonian in Eq.~\eqref{eq:Hamiltonian-wg-3ring-single-particle}. As one can verify, 
the leading order behavior is determined by the LSDs where the most vertices are covered while the least $\epsilon$-edges are involved. 
With Eqs.~\eqref{eq:kappa_alpha_i} and \eqref{eq:p_alpha_i_special} the eigenvalue branches in leading order are given by
\begin{eqnarray}
    \lambda_m (\epsilon_1) & \approx & (A^4 R \epsilon_1 e^{2 \pi i m})^{\frac{1}{6}}, \hspace{15pt} m = 1, \ldots, 6 \label{eq:3ring_tp1} \\
    \lambda_m (\epsilon_2) & \approx & \left\{ \begin{array}{l l} 
    (A^2 R \epsilon_2 e^{2 \pi i m})^{\frac{1}{4}}, & m=1,\ldots,4 \\
    0, & m = 5, 6
    \end{array} \right. \label{eq:3ring_tp2} \\
    \lambda_m (\epsilon_3) & \approx & \left\{ \begin{array}{l l} 
    (R \epsilon_3 e^{2 \pi i m})^{\frac{1}{2}}, & \hspace{12pt} m=1,2 \\
    0, & \hspace{12pt} m = 3, \ldots, 6
    \end{array} \right. . \label{eq:3ring_tp3}
\end{eqnarray}

In Ref.~\cite{2023_kullig}, a cavity-selective scaling of the eigenvalues of such a system is observed. Depending on near which cavity the particle is placed, the eigenvalue splitting 
\begin{equation}
    \Delta \lambda := \max_{i, j} | \lambda_i - \lambda_j | \label{eq:eig_split}
\end{equation}
is proportional to $|\epsilon_1|^{1/6}$ for a particle at the left, $|\epsilon_2|^{1/4}$ at the middle, and $|\epsilon_3|^{1/2}$ at the right cavity. 
Here, the leading order power governs this effect which can be seen in Eqs.~\eqref{eq:3ring_tp1}-\eqref{eq:3ring_tp3}.
From a graph perspective it can be traced back to the number of vertices covered, concretely 6, 4, and 2, by the relevant LSDs shown in Fig.~\ref{fig:lin_sub_digraphs_3ring-wg_single-particle}(b), (d), and (f). By taking the inverse of the concrete numbers which complies with Eq.~\eqref{eq:kappa_alpha_i}, we obtain the leading order powers of $1/6$, $1/4$, and $1/2$. 

The graph perspective leads one to the travelling light wave interpretation. Due to the perturbative backscattering introduced by the single particle, there are cyclic light paths in the system possible. For a particle near the left cavity, such a path can be seen as light travelling from the left via the middle to the right cavity and then reflected by the mirror with the same path reversed; backscattering introduced by the particle at the left cavity leads to a repetition of the path. In an analog fashion, one can imagine the cyclic paths for a particle near the middle or the right cavity. Cavities to the left of the microring with the particle are not part of the cyclic path. Within this interpretation, these paths are responsible for the leading-order behavior of the Puiseux series. 

Another important insight is the number of eigenvalue branches which split away from the EP, see Fig.~\ref{fig:lin_sub_digraphs_3ring-wg_single-particle}(c), (e), and (g). It is also governed by the number of covered vertices in the graph. 
Potentially, this effect can be used to detect at which cavity the particle is present simply by counting the peaks in the reflection spectra measured along the waveguide. 

\subsection{Three waveguide-coupled microrings: Approximation two-parameter perturbation} 
\label{sec:multi_parameter}

Before concluding this paper, let us discuss one last point related to the setup consisting of the 3 waveguide-coupled microrings from the previous \cref{sec:single_particle_perturbation}.
In Ref.~\cite{2023_kullig}, a saturation effect of the eigenvalue splitting in Eq.~\eqref{eq:eig_split}
for single-particle perturbations is observed within finite element method (FEM) simulations of the dielectric structure. The splitting becomes approximately constant below a threshold $\epsilon^{\text{crit}}_i$ for the absolute perturbation parameter $\epsilon_i$ of a single-particle perturbation. 
This effect can not be described with the Hamiltonian used in \cref{sec:single_particle_perturbation}. 
The statements from previous sections about the waveguide-coupled microring system neglect the occurrence of backscattering of light within the waveguide-microring coupling as well as backscattering due to the surface roughness of the microrings. 

In order to describe the saturation effect with an effective Hamiltonian, one needs to take two perturbation causes into account: the single-particle perturbation---which we have already taken into account so far---and additionally also the perturbation induced by surface roughness and waveguide backscattering. 
The latter can be modelled with the perturbation matrix
\begin{equation}
    H_1^{D}  =  \left( \begin{array}{c c c c c c}
        0 & 0 & 0 & 0 & 0 & 1 \\
        0 & 0 & 0 & 0 & 1 & 0 \\
        0 & 0 & 0 & 1 & 0 & 0 \\
        0 & 0 & 1 & 0 & 0 & 0 \\
        0 & 1 & 0 & 0 & 0 & 0 \\
        1 & 0 & 0 & 0 & 0 & 0
    \end{array} \right) . \label{eq:H1_D_wg-3ring}
\end{equation}
The full Hamiltonian 
\begin{equation}
    H = H_0^{\text{EP}_6} + \epsilon_i H_1^{\text{tp}_i} + \epsilon_D H_1^{D} \,,\label{eq:H_multi_par}
\end{equation}
is described by the unperturbed part $H_0^{\text{EP}_6}$ in Eq.~\eqref{eq:H0_wg-3ring}, the single-particle perturbation matrix $H_1^{\text{tp}_i}$ in Eqs.~\eqref{eq:H1_tp1_wg-3ring}-\eqref{eq:H1_tp3_wg-3ring} with $i = 1,2, 3$, and the newly introduced perturbation matrix in Eq.~\eqref{eq:H1_D_wg-3ring}. We have to deal with a multiparameter perturbation since two distinct perturbation parameters, $\epsilon_i$ and $\epsilon_D$, are present.

\begin{figure}[!tb]
	\begin{center}
		\includegraphics[width=1.0\columnwidth]{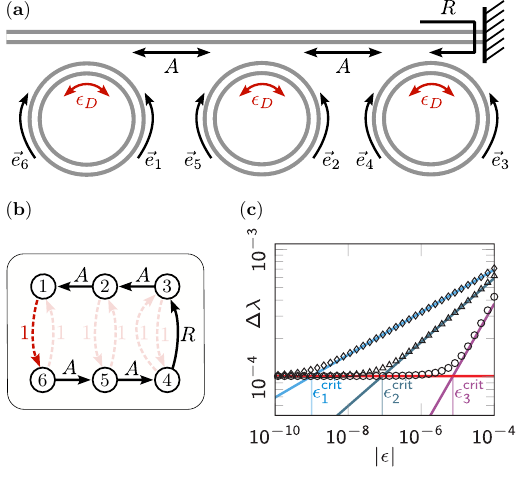}
\caption{(a) Schematic view of the three waveguide-coupled microrings with mirror-induced asymmetric backscattering from Ref.~\cite{2023_kullig} where the parameters $A$ and $ R$ of the Hamiltonian~$H$ are indicated. The perturbation induced by backscattering at the waveguide-microring interface and surface roughness is modelled by the parameter~$\epsilon_D$. The basis of~$H$ is chosen as the traveling waves indicated by the arrows~$\vec{e}_{i}$ with $i = 1,2, \ldots,6$. (b) The relevant LSD (highlighted) for the leading order in the Puiseux series. Toned down and highlighted edges form the graph representation of~$H$. (c) The dimensionless eigenvalue splitting  in Eq.~\eqref{eq:eig_split} over the absolute perturbation strength $|\epsilon|$ induced by a single particle. The red horizontal line show the case where only the $\epsilon_D$-perturbation is present. The other three colored curves show the case where only a single-particle perturbation is present at the left (blue), middle (green), and right (purple) cavity. Unfilled markers show $\Delta \lambda$ where the eigenvalues are numerically obtained from~$H$ where the $\epsilon_D$- and $\epsilon_i$-perturbations with $i = 1,2,3$ are considered. The diamond, triangle, and circle marker correspond the cases where the particle is placed near the left, middle, and right cavity, respectively. The data shown is obtained by pure real values of $\epsilon_i$, ranging from $10^{-10}$ to $10^{-4}$, $A = 4.7 \cdot 10^{-4}$, $R = 3.5 \cdot 10^{-4}$, and $\epsilon_D = 10^{-9}$.
}
		\label{fig:lin_sub_digraphs_3ring-wg_internal-backscattering}
	\end{center}
\end{figure}

The unfilled markers with the black frame in Fig.~\ref{fig:lin_sub_digraphs_3ring-wg_internal-backscattering}(c) show the eigenvalue splitting, Eq.~\eqref{eq:eig_split}, obtained by numerically diagonalizing the effective Hamiltonian in Eq.~\eqref{eq:H_multi_par} where both types of perturbations are present. 
For values of $|\epsilon|$ greater than $\epsilon_{i}^{\text{crit}}$ the characteristic single-particle behavior, where the frequency splitting goes with the 6-th, 4-th, or 2-nd root over $|\epsilon|$ for a particle at the left (diamond marker), middle (triangle marker), or right (circle marker) microring, becomes visible.
The before described saturation effect of the frequency splitting for the absolute single-particle perturbation strength $|\epsilon|$ for values smaller than $\epsilon_{i}^{\text{crit}}$ is present for particles at the left, middle, and right cavity.

To better understand this behavior, let us go a step back and consider both perturbations separately in order to characterize the saturation effect with the graph-theoretical approach.
We start with the additional perturbation induced by surface roughness and waveguide backscattering.
In Fig.~\ref{fig:lin_sub_digraphs_3ring-wg_internal-backscattering}(a), we schematically show how the perturbation is modelled within the effective Hamiltonian framework. 
The perturbation introduces a coupling between the CW and CCW waves of each microring, visualized with the red arrows with the coupling parameter~$\epsilon_D$. 

The corresponding graph representation is depicted in Fig.~\ref{fig:lin_sub_digraphs_3ring-wg_internal-backscattering}(b). Herein the LSD with the most vertices covered while the least $\epsilon_D$-edges involved, which determines the leading-order behavior, is highlighted. 
This LSD covers all six vertices, while one edge related to the perturbation is present. 
With Eqs.~\eqref{eq:kappa_alpha_i} and~\eqref{eq:p_alpha_i_special}, we obtain the Puiseux series in leading order as
\begin{equation}
    \lambda_m (\epsilon_D) = (A^4 R \epsilon_D e^{2 \pi i m} )^{\frac{1}{6}}, \; m = 1, \ldots, 6 . \label{eq:lambda_D}
\end{equation}
This equation is the same as Eq.~\eqref{eq:3ring_tp1}, which comes from a particle perturbation in vicinity of the left microring.
Thus, we see that a plot of the eigenvalues [Eq.~\eqref{eq:lambda_D}] in asymptotically leading order would be---up to a complex phase---identical to Fig.~\ref{fig:lin_sub_digraphs_3ring-wg_single-particle}(c), which depicts the eigenvalues given in Fig.~\ref{fig:lin_sub_digraphs_3ring-wg_single-particle}(c).

We note that this type of perturbation can be characterized as generic due to the sixth-root behavior. 
As a result, for the system consisting of waveguide-coupled microrings with mirror-induced asymmetric backscattering, even without a single-particle perturbation the system is not exactly at an EP.
This can also be seen in Fig.~\ref{fig:lin_sub_digraphs_3ring-wg_internal-backscattering}(c), where the red line corresponds to the eigenvalue splitting obtained with the eigenvalues of the pure $\epsilon_D$-perturbation by setting $\epsilon_i = 0$ and $\epsilon_D = 10^{-9}$ in Eq.~\eqref{eq:H_multi_par}.

Equipped with the above, we can now derive an expression for the saturation value of the eigenvalue splitting [Eq.~\eqref{eq:eig_split}].
Specifically, using Eq.~\eqref{eq:lambda_D}, one obtains
\begin{equation}
    \Delta \lambda^{\text{sat}} = 2 \cdot | A^4 R \epsilon_D |^{\frac{1}{6}} \label{eq:lamb_sat} \,.
\end{equation}
The saturation value is governed by the perturbation induced by surface roughness and waveguide backscattering.
Note that the generic sixth-root behavior can lead to a considerable eigenvalue splitting even for small perturbations strength $\epsilon_D$.

Next, we consider the case where only the single-particle perturbations are present.
This case was already discussed in the previous \cref{sec:single_particle_perturbation}; here we only repeat the results.
The lines with the colors blue, green, and purple in Fig.~\ref{fig:lin_sub_digraphs_3ring-wg_internal-backscattering}(c) are determined by the eigenvalue splitting, Eq.~\eqref{eq:eig_split}, of the pure single-particle perturbations in the Eqs.~\eqref{eq:3ring_tp1}-\eqref{eq:3ring_tp3}, respectively. 
Their slopes are given by the leading order power of $1/6$, $1/4$, and $1/2$ which can be easily obtained by the covered vertices of the highlighted LSDs shown in \cref{fig:lin_sub_digraphs_3ring-wg_single-particle}(b), (d), and (f), respectively. 
For these lines, the effective Hamiltonian is given by setting $\epsilon_D$ in Eq.~\eqref{eq:H_multi_par} to zero. It is the same as in \cref{sec:single_particle_perturbation}.

Let us now bridge the separately considered perturbations to describe the saturation effect.
For particle sizes where the related absolute perturbation parameter strength $| \epsilon_i |$ is much bigger than the critical value $\epsilon_i^{\text{crit}}$, the behavior related to the single-particle perturbation becomes dominant, as one can see in Fig.~\ref{fig:lin_sub_digraphs_3ring-wg_internal-backscattering}(c) via a comparison with the unfilled markers (numerical results include both types of perturbations). 
The critical threshold value $\epsilon^{\text{crit}}_i$ is obtained by the intersection of the red line corresponding to the pure $\epsilon_D$-perturbation with the blue, green, or purple line related to the pure $\epsilon_i$-perturbation in Fig.~\ref{fig:lin_sub_digraphs_3ring-wg_internal-backscattering}(c).
It can be found formally by setting the absolute value of Eq.~\eqref{eq:lambda_D} equal to the absolute value of the non-vanishing branches of Eqs.~\eqref{eq:3ring_tp1}-\eqref{eq:3ring_tp3}, respectively, and solving for the absolute value of $\epsilon_i$ with $i=1,2,3$ which leads to 
\begin{eqnarray}
    \epsilon_1^{\text{crit}} & = & | \epsilon_D | , \label{eq:eps_crit1} \\
    \epsilon_2^{\text{crit}} & = & \frac{(|A| |\epsilon_D|)^{\frac{2}{3}}}{|R|^{\frac{1}{3}}} , \\
    \epsilon_3^{\text{crit}} & = & \frac{|A|^{\frac{4}{3}}}{|R|^{\frac{2}{3}}} | \epsilon_D |^{\frac{1}{3}} . \label{eq:eps_crit3}
\end{eqnarray}
The absolute perturbation strength $|\epsilon_D|$ has to be sufficiently small in a way that the system is still in close vicinity to the EP parameter configuration.  

To summarize, by separately considering the single-particle perturbation at the one hand and the perturbation related to waveguide-backscattering and surface roughness at the other hand,
the saturation effect for the eigenvalue splitting can be approximated. 
Our graph-theoretical approach delivers a convenient tool to extract the relevant information to do so. 
The saturation value is stated in Eq.~\eqref{eq:lamb_sat} and the critical perturbation strength of the single particle, for which a transition between saturation and single-particle behavior is marked, is approximated in the Eqs.~\eqref{eq:eps_crit1}-\eqref{eq:eps_crit3}. 

For the general case of a multi-parameter perturbed Hamiltonian like in Eq.~\eqref{eq:H_multi_par}, our presented approach reaches its limits.
It is not set up to resolve the higher-dimensional parameter space introduced by the additional perturbation parameters.  
However, in Ref.~\cite{2003_beringer} is an algorithm for the determination of the Puiseux series in multiple variables described. It uses a generalization of the Newton polygon into higher dimensions, the so-called Newton polyhedron. 
This can be seen as the equivalent of the determination algorithm described in Ref.~\cite{1950_walker} which is used for our approach generalized to a multi-parameter perturbation scenario. 
In order to connect this to graph theory, one can proceed in a similar fashion as in \cref{sec:graph_approach}. For instance, one would graph-theoretically determine the coefficients of the characteristic polynomial $b_{\mu, \gamma, \nu}$ of the Hamiltonian in Eq.~\eqref{eq:H_multi_par} in the variables $\epsilon_i, \epsilon_D$ and the eigenvalues $\lambda$. See \cref{app:multi_par} for a description of this ansatz. 

\section{Conclusion}

The eigenvalue spectra of systems with an EP of order~$n$ under perturbations with strength $\epsilon$ exhibit a wide variety of characteristic behaviors. Perturbations are generic if the eigenvalues scale proportional to $\epsilon^{1/n}$ for small $\epsilon$. The generic case is usually desired for sensor applications due to the strong response of the eigenvalues with respect to weak perturbations.
An eigenvalue behavior that is different from the generic case is called non-generic. With the order $n$ of the EP the varieties in the eigenvalue behavior increase for non-generic perturbations.

In order to characterize the spectrum of EPs under linear perturbations in a basis-independent way, with respect to the (effective) Hamiltonian, we used the coefficients of the characteristic polynomial. We derived a combinatorial formula to calculate these coefficients for the characteristic polynomial in the eigenvalues and one additional parameter, here the perturbation strength, based on graph theory by considering linear sub digraphs (LSDs) that are cyclic configurations of the graph representations of the Hamiltonian.
Our approach combines graph theory with the Puiseux series to determine the behavior of the eigenvalue spectrum of perturbed higher-order EPs. 
The asymptotically leading-order behavior of the eigenvalues is determined by the LSDs where the most vertices of the graph picture are covered while the least edges corresponding to a perturbation are involved. 

The approach is applied to coupled microring systems with a higher-order EP. Via different types of perturbations the relevance of understanding the non-generic behavior is highlighted. The considered perturbations are not exotic in a sense that one has to tailor them in order to see the desired effect. The heating of a microring, placing a particle near a microring, and backscattering/surface roughness effects are investigated. With the graph-theoretical approach one can interpret for such systems that the reasoning for the eigenvalue behavior are configurations of specific cyclic paths which the light can travel in the system.  

An extension of our graph-theoretical approach to multi-parameter perturbations is sketched and motivated with the saturation effect occurring for the cavity-selective sensing of waveguide-coupled microrings with mirror-induced asymmetric backscattering, observed in Ref.~\cite{2023_kullig}. 
Here the perturbation introduced by the backscattering at the waveguide-microring coupling and surface roughness leads to a saturation effect of the frequency splitting for a single-particle perturbation strength smaller than a certain critical value.

We will motivate another potentially useful case of the presented approach. It can become a helpful tool for inventing integrated devices based on higher-order EPs. Typically the components of an integrated device are via physical mechanisms coupled among each other which can be expressed naturally with edges in the graph picture. By modelling the perturbation within this framework one can directly deduce the perturbative eigenvalue behavior and adapt the design according to the desired features of the device.

Future directions for the graph-theoretical perspective to non-Hermitian systems with EPs could be the exploration of eigenvalue braids and knots via the mathematically well-known concepts arising in the field of plane algebraic curves \cite{2012_brieskorn_alg_curves}, the extension to general multi-parameter perturbations, and the study of non-Hermitian symmetries~\cite{MB21,SK22}, where non-generic perturbations can appear naturally. 

\acknowledgments 
We thank the DFG (Grant No.~WI 1986/14-1) for financial support. We acknowledge support for the Book Processing Charge by the Open Access Publication Fund of Magdeburg University.

\clearpage

\appendix

\section{On combinatorics} \label{app:char_poly}

The combinatorial formula in \cref{eq:char_poly_combi_main} of the characteristic polynomial is a special case of the formula of the determinant of the sum of two square matrices $A$ and $B$ with dimension $n$ \cite{1990Marcus}, which reads
\begin{eqnarray}
    \det{(A + B)} & = & \sum_{p = 0}^n \sum_{\substack{\alpha, \beta \\\in K_{n, p}}}  (-1)^{s(\alpha) + s(\beta)} \times \nonumber \\
    & & \hspace{0pt} \text{det} \bm{(} A[\alpha | \beta] \bm{)} \text{det} \bm{(} B(\alpha | \beta) \bm{)}\label{eq:det_sum}
\end{eqnarray}
where the set $K_{n,p} = \{ (i_1, \ldots, i_p) \in \mathbb{N}^p | 1 \leq i_1 < \ldots< i_p \leq n \}$ describes the combinations without repetition and without exchange (k-combination) of $i_1, \ldots, i_p$.
The square submatrix $A[\alpha|\beta]$ (square brackets) of dimension~$p$ is given by considering only rows with indices appearing in $\alpha \in K_{n, p}$ and columns appearing in $\beta \in K_{n, p}$.
The concept of induced subgraphs introduced in \cref{sec:graph_foundation} is a one-to-one-mapping between a submatrix $A[\alpha | \alpha] = A[\alpha]$ and its graph representation. 
The square submatrix $B(\alpha|\beta)$ (round brackets) of dimension $n-p$ is given by the remaining parts of the matrix $B$ after deleting the rows corresponding to $\alpha$ and the columns corresponding to $\beta$; this operation can also be mapped to a graph representation.

The sum of the integers in $\alpha, \beta$ is denoted with $s(\alpha)$, $s(\beta)$, respectively. For $p = 0$, the summand means $\det (B)$ and for $p = n$, it is $\det (A)$.

For fluent reading, we state again the combinatorial formula in \cref{eq:char_poly_combi_main} for the characteristic polynomial of a generic matrix $A \in \mathbb{C}^{n \times n}$  \cite{1990Marcus, Brualdi2008CombinatorialApproachMatrixTheory} 
\begin{equation}
	\det{(\lambda \mathbb{1} - A)} = \sum_{p = 0}^n (-1)^{n+p} \lambda^p c_{n-p} (A)
	\label{eq:char_poly} \,,
\end{equation} 
where $c_{n-p}(A)$ equals the sum of all principal minors of order $n-p$ of $A$. A principal minor is defined as the determinant of a submatrix. 
In equation form, this reads
\begin{equation}
	c_{n-p} (A) = \sum_{\substack{\alpha \in K_{n, p}}} \det{\bm{(}A(\alpha | \alpha)\bm{)}} \label{eq:cnp} \,,
\end{equation}
where $A(\alpha | \alpha)$ has the dimension (order) $n-p$.
Note that $c_n = \det{A}$, $c_1 = \text{trace}\,A$, and $c_0 := 1$.

We consider a matrix $H = H_0 + \epsilon H_1$. From Eq.~\eqref{eq:cnp}, we have a look at an arbitrary term which is a principal minor. In other words, the determinant of a matrix $H(\gamma | \gamma) = H_0(\gamma | \gamma) + \epsilon H_1(\gamma | \gamma)$ with dimension $n-p$ and $\gamma \in K_{n,p}$. With Eq.~\eqref{eq:det_sum} we obtain
\begin{eqnarray}
    \det \tilde{H} & = & \sum_{j = 0}^{n-p} \sum_{\substack{\alpha, \beta \\\in K_{n-p,j}}} (-1)^{s(\alpha) + s(\beta)} \times \nonumber \\
    & & \epsilon^j \det \bm{(}\tilde{H}_1[\alpha|\beta] \bm{)} \det \bm{(}\tilde{H}_0 (\alpha|\beta) \bm{)} \label{eq:det_sum_H}
\end{eqnarray}
where $\tilde{H} = H(\gamma | \gamma)$, $\tilde{H}_0 = H_0(\gamma | \gamma)$, and $\tilde{H}_1 = H_1(\gamma | \gamma)$. Equation~\eqref{eq:det_sum_H} tells us that only integer powers of $\epsilon$ arise in the terms of Eq.~\eqref{eq:cnp} where the maximum power in $\epsilon$ is given by $n-p$ if the corresponding $\det (\tilde{H}_1[\alpha|\beta]) \neq 0$ and $\det (\tilde{H}_0 (\alpha|\beta)) \neq 0$.
The summand for $j = 0$ is simply given by $\det\tilde{H}_0$.

With these insights, we can rewrite the characteristic polynomial in Eq.~\eqref{eq:char_poly} in $\epsilon$ and $\lambda$ for $H$ 
\begin{eqnarray}
	\text{det } (\lambda \mathbb{1} - H) & = & \sum_{\nu = 0}^{n} \sum_{\mu = 0}^{n - \nu}  a_{\mu, \nu} \epsilon^{\mu} \lambda^{\nu}  \label{eq:char_poly_sorted} \\
 \text{with } a_{\mu, \nu} & = & (-1)^{n + \nu} \frac{1}{\mu !} \frac{\partial^{\mu} c_{n-\nu} (H)}{\partial \epsilon^{\mu}} \bigg|_{\epsilon = 0} . \label{eq:a_mu_nu_0}
\end{eqnarray}
The $\mu$-th derivative with respect to $\epsilon$ at $\epsilon = 0$ in Eq.~\eqref{eq:a_mu_nu_0} in combination with the factor $1/\mu !$ gives us the coefficients of $\epsilon^{\mu}$ arising in Eq.~\eqref{eq:cnp}. 
Note the slightly different notation  in Eq.~\eqref{eq:char_poly_sorted} if compared to \cref{eq:cPs} in the main part. For the appendix, we stick to the notation in Eq.~\eqref{eq:char_poly_sorted}, which is commonly used in the literature. At the end of this section, the link between the two notations is described. 

We consider the summand for $\mu = 0$ in Eq.~\eqref{eq:char_poly_sorted}
\begin{equation}
	\sum_{\nu = 0}^n a_{0, \nu} \lambda^{\nu}
 = \det{(\lambda \mathbb{1} - H_0)} \label{eq:zero_a0nu}
\end{equation}
which equals the characteristic polynomial of $H_0$. 
We demand that all $n$ eigenvalues of $H_0$ are equal to zero. This leads to the characteristic polynomial 
\begin{equation}
	\text{det }(\lambda \mathbb{1} - H_0) =  \lambda^n . \label{eq:char_H0}
\end{equation}

A comparison of the coefficients in Eqs.~\eqref{eq:zero_a0nu} and \eqref{eq:char_H0} reveals that 
\begin{equation}
	a_{0, \nu} = \left\{ 
 \begin{array}{c c} 
 1 & \text{for } \nu = n \\
 0 & \text{else}
 \end{array} \right.
 . \label{eq:a_0_nu}
\end{equation}

The requirement that all $n$ eigenvalues of $H_0$ are zero is fulfilled, for example, when $H_0$ is nilpotent of order $n$ or if $H_0$ has a block diagonal structure with each block being nilpotent with lower order. 
As mentioned in the main part, this translates to the scenarios with one EP of order $n$ or multiple EPs possibly of different order but with the same degenerate eigenvalue.

\begin{figure}[htb]
	\begin{center}
	\includegraphics[width=\columnwidth]{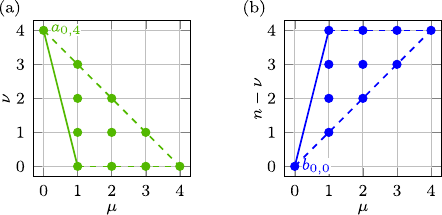}
	\caption{Visualization of the carrier for the characteristic polynomial in Eq.~\eqref{eq:char_poly_sorted_simp0} with $n=4$, $a_{\mu, \nu} \neq 0$ and the corresponding Newton polygon (a) in the sense of the coefficients $a_{\mu, \nu}$. (b) Mapped to a coordinate system with $b_{\mu, v} = a_{\mu, n-v}$ which is more suited for graph-theoretical considerations.
 } 
	\label{fig:carrier_generic}
	\end{center}
\end{figure}

Through consideration of Eq.~\eqref{eq:a_0_nu} in Eq.~\eqref{eq:char_poly_sorted}, we obtain
\begin{equation}
    \det (\lambda \mathbb{1} - H) = \lambda^n + \sum_{\nu = 0}^{n-1} \sum_{\mu = 1}^{n - \nu}  a_{\mu, \nu} \epsilon^{\mu} \lambda^{\nu} . \label{eq:char_poly_sorted_simp0} \\
\end{equation}
This characteristic polynomial has a special structure, which can be seen with a visualization of the corresponding carrier $\{ (\mu, \nu) | a_{\mu, \nu} \neq 0 \}$ of Eq.~\eqref{eq:char_poly_sorted_simp0}~ as exemplary shown in \cref{fig:carrier_generic}(a) for $n = 4$.
Note that $a_{0, 4} = 1$ and the remaining $a_{\mu, \nu}$ in Eq.~\eqref{eq:char_poly_sorted_simp0} are dependent on the matrix~$H$ at hand. Some of them could potentially also vanish. 
Because of the property described by Eq.~\eqref{eq:a_0_nu}, the existence of a Puiseux series is guaranteed through Theorem 3.2 and Theorem 3.3 of Ref.~\cite{1950_walker}.

So far, we have shown how the structure of the characteristic polynomial of a higher-order EP or multiple EPs with the same eigenvalue, namely zero, under linear perturbation in $\epsilon$ and $\lambda$ looks like. 
We remark that an EP with any degenerate eigenvalue can be spectrally shifted. Hence, the prerequisite for the degenerate eigenvalue of zero is no restriction in generality. 

Next, we introduce the graph-theoretical perspective with the Coates determinant formula in Eq.~\eqref{eq:coates_det_main} and state it again for convenience~\cite{Brualdi2008CombinatorialApproachMatrixTheory}
\begin{equation*}
    \det{A} = \sum_{L \in \mathcal{L} (A)} (-1)^{n - c (L)} w (L)
\end{equation*}
where the set of all LSDs of $A\in \mathbb{C}^{n \times n}$ is $\mathcal{L} (A)$. 
The number of disconnected cycles of a LSD $L$ is given by $c(L)$ and the product of the edge weights is called $w(L)$.
In Ref.~\cite{Brualdi2008CombinatorialApproachMatrixTheory}, the underlying graph representation of the matrix $A$ is called a Coates digraph, where only a single edge pointing from vertex~$j$ towards vertex~$i$ is allowed.

In order to reformulate the coefficients of the characteristic polynomial given in Eq.~\eqref{eq:a_mu_nu_0}, we use Eq.~\eqref{eq:coates_det_main} for the principal minors appearing in Eq.~\eqref{eq:cnp} and obtain
\begin{equation}
    a_{\mu, \nu} = \sum_{\alpha \in K_{n, \nu}} \sum_{L \in \mathcal{L} (H(\alpha | \alpha))} (-1)^{c(L)} \frac{1}{\mu !} \frac{\partial^{\mu} w(L)}{\partial \epsilon^{\mu}} \bigg|_{\epsilon = 0} . \label{eq:a_mu_nu_graph0}
\end{equation}
The first sum in Eq.~\eqref{eq:a_mu_nu_graph0} determines the indices $\alpha$ for all possible submatrices $H(\alpha | \alpha)$ of dimension $n - \nu$. The second sum goes over all possible LSDs $L$ of such a submatrix $H(\alpha | \alpha)$. The sign of the summand is determined by the number of cycles of $L$. It is positive for an even number, and negative for an odd number. 

Let us now reformulate \cref{eq:a_mu_nu_graph0}.
The basic idea is as follows: In \cref{eq:a_mu_nu_graph0}, we evaluate all possible LSDs, compute their respective product of edge weights and then select, by means of the partial derivative, the terms where the exponent of $\epsilon$ is equal to $\mu$.
This process is rather inefficient; a better way is to extend the concept of LSDs by allowing them to operate also on multi digraphs.
We introduced these in the main text; they are graphs where two edges from any vertex~$i$ towards vertex~$j$ are possible; we needed them to distinct edges related to~$H_0$ and~$H_1$.
To make full use of multi digraphs, we introduce the set $\mathcal{L}_{\mu, \nu} (H)$ which contains all LSDs of the submatrices of order $\nu$ where $\mu$ times an $\epsilon$-edge is involved. Or in other words, LSDs which cover $\nu$ vertices while $\mu$ times an $\epsilon$-edge is involved.
Equipped with this new concept, it can be easily shown that \footnote{To see this, let us take a matrix of the form $M = A + \epsilon B$.
It can be represented either as a multi digraph (with two types of edges: Those coming from A, and those coming from $B$), or as a Coates digraph with only one type of edge, but with the edge weight containing information of both $A$ and $\epsilon B$. 
Let us start with the latter; that is, we represent $M$ as a Coates digraph.
To show the underlying logic, assume that there is only one LSD containing two vertices, with respective weights $a_1 + \epsilon b_1$, and $a_2$.
The product of weights is thus $a_1 a_2 + \epsilon a_2 b_1$.
If we are only interested in the prefactor of the term proportional to $\epsilon$, that is, in $a_2 b_1$.
One way would be to extract it using the partial derivative, as done in \cref{eq:a_mu_nu_graph0}.
Alternatively, we could represent $M$ as a multi digraph, and search for all LSDs that contain two vertices and one $\epsilon$-edge; there would only be one such graph, with product of weights given by $a_2 b_1$. 
By extending this logic---that is, using the distributive property of the product of weights---, one directly sees that \cref{eq:a_mu_nu_graph0} and \cref{eq:deriv_a_mu_nu} are equivalent.} \cref{eq:a_mu_nu_graph0} is equivalent to
\begin{equation}
    a_{\mu, \nu} = \sum_{L \in \mathcal{L}_{\mu, n-\nu} (H)} (-1)^{c(L)} w(L) \big|_{\epsilon = 1} . \label{eq:deriv_a_mu_nu}
\end{equation}
We remark that \cref{eq:a_mu_nu_graph0,eq:deriv_a_mu_nu} holds even for a general $H_0$. There is no need for $H_0$ that all eigenvalues have to be zero. As long as one is interested in the coefficients of the characteristic polynomial of a matrix of the form $H = H_0 + \epsilon H_1$, \cref{eq:deriv_a_mu_nu} is valid. 
We can apply the same logic to reformulate Eq.~\eqref{eq:char_poly} 
\begin{equation}
    \det ( \mathbb{1} - A ) = \sum_{\nu = 0}^{n} a_{\nu} \lambda^\nu
\end{equation}
in terms of the coefficients of the characteristic polynomial
\begin{equation}
    a_{\nu} = \sum_{L \in \mathcal{L}_{n-\nu} (H)} (-1)^{c(L)} w(L) .
\end{equation}
The set $\mathcal{L}_{n-\nu} (A)$ contains all LSDs of the submatrices of order $n-\nu$.
This logic was used in Ref.~\cite{Brualdi2008CombinatorialApproachMatrixTheory} to derive Eq.~\eqref{eq:char_poly}.  

To summarize our achievements. We were able to introduce a multi digraph which allows one to combinatorically determine the coefficients of the characteristic polynomial in the eigenvalues $\lambda$ and one additional parameter $\epsilon$, which is valid for any matrix $H = H_0 + \epsilon H_1$.
We use this result to connect graph theory with the Puiseux series, which is discussed in \cref{app:plane_alg_curves} and \cref{sec:graph_approach}.
We can even expand this concept to matrices of the form $H = H_0 + X_1 H_1 + X_2 H_2 + \cdots$ which is sketched in \cref{app:multi_par}.

In the main part we use the convention 
\begin{equation}
    b_{\mu, \nu} := a_{\mu, n - \nu} \label{eq:connection_b_a}
\end{equation}
for a more convenient graph-theoretical interpretation of the coefficients of the characteristic polynomial. It leads to 
\begin{equation}
    b_{\mu, \nu} = \sum_{L \in \mathcal{L}_{\mu, \nu} (H)} (-1)^{c(L)} w(L) \big|_{\epsilon = 1} .  \label{eq:b_mu_nu_app_A}
\end{equation}
Note that in Eq.~\eqref{eq:b_mu_nu_graph_main} no evaluation for $\epsilon = 1$ appears in contrast to Eq.~\eqref{eq:b_mu_nu_app_A} because we define the graph representation in \cref{sec:graph_quantities} such that no $\epsilon$ appears explicitly in the edge weights. 
The mapping between the coefficients $a$ and $b$ can be seen as shift of all carrier points in \cref{fig:carrier_generic}(a) in negative $\nu$ direction by the value $n$ and then all points are mirrored along the $\mu$-axis to obtain \cref{fig:carrier_generic}(b).

\section{On plane algebraic curves} \label{app:plane_alg_curves}

The research of plane algebraic curves considers polynomials in two variables 
\begin{equation}
    f (X, Y) = \sum_{\mu, \nu} a_{\mu \nu} X^{\mu} Y^{\nu} \,,\label{eq:poly_2var}
\end{equation}
with complex-valued coefficients $a_{\mu \nu}$ which is in our case the characteristic polynomial in Eq.~\eqref{eq:char_poly_sorted_simp0} where the two variables $\epsilon \rightarrow X$ and $\lambda \rightarrow Y$.
As elaborated in the main part, the roots of \cref{eq:poly_2var}, namely
\begin{equation}
    f (X, Y) = 0 \label{eq:plane_alg_curve},
\end{equation}
can be seen as a geometric structure, the plane algebraic curve.
One method to characterize these curves is to parameterize them (locally) in a way that
\begin{equation}
    f(X, \varphi (X)) = 0 \label{eq:pui_converge_def}
\end{equation}
where
$\varphi (X)$
is a convergent Puiseux series in fractional powers of $X$
with the structure given in Eq.~\eqref{eq:puiseux_struc}. 
The reason why such a Puiseux series for our purpose exists lies in Theorem~3.2 and Theorem~3.3 of Ref.~\cite{1950_walker}. Due to the special structure of the characteristic polynomial in Eq.~\eqref{eq:char_poly_sorted_simp0} the prerequisite of Theorem~3.3 is fulfilled as mentioned in \cref{app:char_poly}. Theorem~3.2 justifies the existence of the eigenvalue branches. 

In order to justify Eqs.~\eqref{eq:kappa_alpha_i} and \eqref{eq:p_alpha_i}, the construction scheme of the Puiseux series described in Ref.~\cite{1950_walker} is used. 
The idea is to write the Puiseux series in the form
\begin{eqnarray}
    \varphi & = & c_1 X^{\gamma_1} + c_2 X^{\gamma_1 + \gamma_2} + c_3 X^{\gamma_1 + \gamma_2 + \gamma_3} + \ldots \nonumber \\
    & = & X^{\gamma_1} (c_1 + \varphi_1) \label{eq:varphi_0}
\end{eqnarray}
with $c_i \neq 0$, $\gamma_2 > 0$, $\gamma_3 > 0$, $\ldots$ and $\varphi_1 = c_2 X^{\gamma_2} + \ldots$. The $\gamma_1$ and $c_1$ correspond to the leading order of the Puiseux series, equivalent to $\kappa_{\alpha}$ and $p_{\alpha}$ in Eq.~\eqref{eq:puiseux_leading}; $\varphi_1$ contains the higher-order terms of the Puiseux series.
Inserting Eq.~\eqref{eq:varphi_0} into the power series in Eq.~\eqref{eq:poly_2var} gives
\begin{eqnarray}
    f (X, \varphi) & = & \sum_{\nu = 0}^n \sum_{\mu} a_{\mu, \nu} X^{\mu + \nu \gamma_1} (c_1 + \varphi_1)^{\nu} \nonumber \\
    & = & \sum_{\nu = 0}^n \sum_{\mu} (c_1)^{\nu} a_{\mu, \nu} X^{\mu + \nu \gamma_1} + g (X, \varphi_1) \label{eq:f_x_varphi}
\end{eqnarray}
where $g$ contains all the terms involving $\varphi_1$.
A necessary condition for Eq.~\eqref{eq:pui_converge_def} to hold is that the leading-order terms in Eq.~\eqref{eq:f_x_varphi} cancel. 
This means at the one hand that the summands with the smallest power in $X$ and without $\varphi_1$ in Eq.~\eqref{eq:f_x_varphi} have to be identified. 
Hence there must exist at least two indices $j,k = 0, 1, \ldots, n$ for which
\begin{equation}
    \alpha_j + j \gamma_1 = \alpha_k + k \gamma_1 \leq \alpha_{\nu} + \nu \gamma_1
\end{equation}
where $\alpha_i = \min \{ \mu | a_{\mu, i} \neq 0 \}$ and $\nu = 0, 1, \ldots, n$.
This condition can be directly translated into the relevant line segments of the Newton polygon.
Namely, that there is a~$\beta_1$ for which all points of the carrier $\Delta f = \{ (\mu, \nu) | a_{\mu, \nu} \neq 0 \}$ lie on or above the line
\begin{equation}
    \nu + \frac{1}{\gamma_1} \mu = \beta_1 . \label{eq:line_segment_poly}
\end{equation}
The slope of the line $\nu (\mu)$ is given by $-1/\gamma_1$. If multiple relevant line segments exist they have different slopes und hence different leading order fractional power $\gamma_1 = \kappa_{\alpha}$. This is exemplary visualized with the solid line in \cref{fig:carrier_generic}(a).
On the other hand, the coefficients of all terms in leading order must cancel, namely
\begin{equation}
    \sum_{\mu, \nu \in \text{Eq.} \,\eqref{eq:line_segment_poly}} a_{\mu, \nu} (c_1)^\nu = 0 . \label{eq:determination_puiseux_coeff_a_mu_nu}
\end{equation}
We solve for $c_1 \neq 0$ to obtain the leading-order coefficient of the Puiseux series for each relevant line segment which is denoted with $p_{\alpha}$ in the main part. 
For a detailed elaboration of the process of finding the Puiseux series, Ref.~\cite{1950_walker} is recommended. 

Like in \cref{app:char_poly}, the mapping between the different conventions of the coefficients of the characteristic polynomial defined in Eq.~\eqref{eq:connection_b_a} leads to the results shown in the main part, namely Eqs.~\eqref{eq:kappa_alpha_i} and \eqref{eq:p_alpha_i}. 
We remark the subtle difference in the determination of the relevant line segments by using the convention with $a_{\mu, \nu}$'s used in the appendix compared to the $b_{\mu, \nu}$'s in main part, see \cref{fig:carrier_generic}(a) and (b), respectively.
The relevant line segments in terms of the $b_{\mu, \nu}$'s are given by the lines of the Newton polygon where all carrier points lie on or below this lines whereas by the convention with the $a_{\mu, \nu}$'s the carrier points lie on or above the Newton polygon lines. 

\section{On multi-parameter perturbations} \label{app:multi_par}

As seen in \cref{sec:multi_parameter}, the Hamiltonian in Eq.~\eqref{eq:H_multi_par} contains two perturbation parameters, namely $\epsilon_i$ and $D$. One can imagine systems where even more than two perturbation parameters are needed to describe the system's behavior. Here, such scenarios are considered as multi-parameter perturbations. 

In order to extend our graph-theoretical approach to multi-parameter perturbations, we sketch in the following a reasonable ansatz. 
For the sake of simplicity we consider a two-parameter perturbation described by the Hamiltonian 
\begin{equation}
    H = H_0 + X_1 H_1 + X_2 H_2 \label{eq:H_X1_X2}
\end{equation}
like in Eq.~\eqref{eq:H_multi_par} with the two perturbation parameters $\epsilon_i \rightarrow X_1$ and $\epsilon_D \rightarrow X_2$.
The resulting characteristic polynomial 
\begin{equation}
    f (X_1, X_2, Y) = \sum_{\mu, \gamma, \nu} a_{\mu, \gamma, \nu} (X_1)^{\mu} (X_2)^{\gamma} Y^{\nu}
\end{equation}
contains only integer powers of the perturbation parameters and the eigenvalues $\lambda \rightarrow Y$. Like in \cref{app:char_poly}, this can be justified with Eq.~\eqref{eq:det_sum}. Therefore, the coefficients
\begin{equation}
    a_{\mu, \gamma, \nu} = (-1)^{n+\nu} \frac{1}{\mu! \gamma !} \frac{\partial^{\mu + \gamma} c_{n-\nu} (H)}{\partial^{\mu} X_1 \partial^{\gamma} X_2} \bigg|_{X_1, X_2 =0} \label{eq:a_mu_nu_0_multi}
\end{equation}
are given in a similar form as in Eq.~\eqref{eq:a_mu_nu_0}.
Again the Coates determinant formula in Eq.~\eqref{eq:coates_det_main} can be used for the principal minors $c_{n-\nu}$ in Eq.~\eqref{eq:a_mu_nu_0_multi}. With a similar argument to justify Eq.~\eqref{eq:deriv_a_mu_nu} we get
\begin{equation}
    a_{\mu, \gamma, \nu} = \sum_{L \in \mathcal{L}_{\mu, \gamma, n-\nu} (H)} (-1)^{c(L)} w(L) \big|_{X_1, X_2 = 1} \ , \label{eq:deriv_a_mu_gamma_nu}
\end{equation}
where the set $\mathcal{L}_{\mu, \gamma, n-\nu} (H)$ contains all LSDs from all submatrices of dimension $n-\nu$ of $H$ where $\mu$-times an edge related to $H_1$ and $\gamma$-times an edge related to $H_2$ is present. 
A similar argumentative structure applies for deriving the coefficients of the characteristic polynomial described by LSDs with more than two perturbation parameters. 

To connect Eq.~\eqref{eq:deriv_a_mu_gamma_nu} to the eigenvalue spectrum of~$H$ given in Eq.~\eqref{eq:H_X1_X2}, the algorithm for determining the Puiseux series for multi-variate polynomials described in Ref.~\cite{2003_beringer} can be used. It exploits the generalized concept of Newton polygons in higher dimensions, namely the Newton polyhedron. This step is comparable with the usage of the algorithm described in Ref.~\cite{1950_walker} for the determination of the Puiseux series for a single perturbation parameter with the help of the Newton polygon which is used in \cref{sec:newton_polygon}.

\end{document}